\documentclass[pra,reprint,notitlepage,superscriptaddress]{revtex4-1}

\usepackage[english]{babel}
\usepackage[latin9]{inputenc}
\usepackage{amsmath,amssymb,amsfonts,bbold,color,graphicx,comment}
\usepackage[colorlinks,citecolor=blue,urlcolor=blue]{hyperref}

\newcommand{\be}{\boldsymbol{\epsilon}}
\newcommand{\bd}{\mathbf{d}}
\newcommand{\br}{\mathbf{r}}

\newcommand{\brho}{\boldsymbol{\rho}}
\newcommand{\Z}{\mathbb{Z}}

\newcommand{\ket}[1] {| #1 \rangle}
\newcommand{\bra}[1] {\langle #1 |}

\newcommand{\ict}{\affiliation{Center for Quantum Physics, University of Innsbruck, Austria}}
\newcommand{\iqoqi}{\affiliation{Institute for Quantum Optics and Quantum Information of the Austrian Academy of Sciences,  Innsbruck, Austria}}
\newcommand{\jila}{\affiliation{JILA, University of Colorado and National Institute of Standards and Technology, and Department of Physics, University of Colorado, Boulder,  CO 80309, USA}}
\newcommand{\ctqm}{\affiliation{Center for Theory of Quantum Matter, University of Colorado, Boulder, CO 80309, USA}}

\begin{document}
%%TC:ignore
\title{Quantum Many-Body Physics with Ultracold Polar Molecules: Nanostructured Potential Barriers and Interactions}
 
\author{Andreas Kruckenhauser}\ict \iqoqi
\author{Lukas M. Sieberer}\ict \iqoqi
\author{Luigi De Marco}\jila
\author{Jun-Ru Li}\jila
\author{Kyle Matsuda}\jila
\author{William G. Tobias}\jila
\author{Giacomo Valtolina}\jila
\author{Jun Ye}\jila
\author{Ana Maria Rey}\jila \ctqm
\author{Mikhail A. Baranov}\ict \iqoqi
\author{Peter Zoller}\ict \iqoqi
 
\date{\today}

\begin{abstract}
  We design dipolar quantum many-body Hamiltonians that will facilitate the
  realization of exotic quantum phases under current experimental conditions
  achieved for polar molecules. The main idea is to modulate both single-body
  potential barriers and two-body dipolar interactions on a spatial scale of
  tens of nanometers to strongly enhance energy scales and, therefore, relax
  temperature requirements for observing new quantum phases of engineered
  many-body systems. We consider and compare two approaches. In the first,
  nanoscale barriers are generated with standing wave optical light fields
  exploiting optical nonlinearities. In the second, static electric field
  gradients in combination with microwave dressing are used to write
  nanostructured spatial patterns on the induced electric dipole moments, and
  thus dipolar interactions. We study the formation of inter-layer and interface
  bound states of molecules in these configurations, and provide detailed
  estimates for binding energies and expected losses for present experimental
  setups. 
\end{abstract}
\maketitle

\section{Introduction}

\begin{figure}[t!!]
  \centering
  \includegraphics[clip,trim=6.9cm 10.3cm 5.8cm 10.5cm,width=0.48\linewidth]{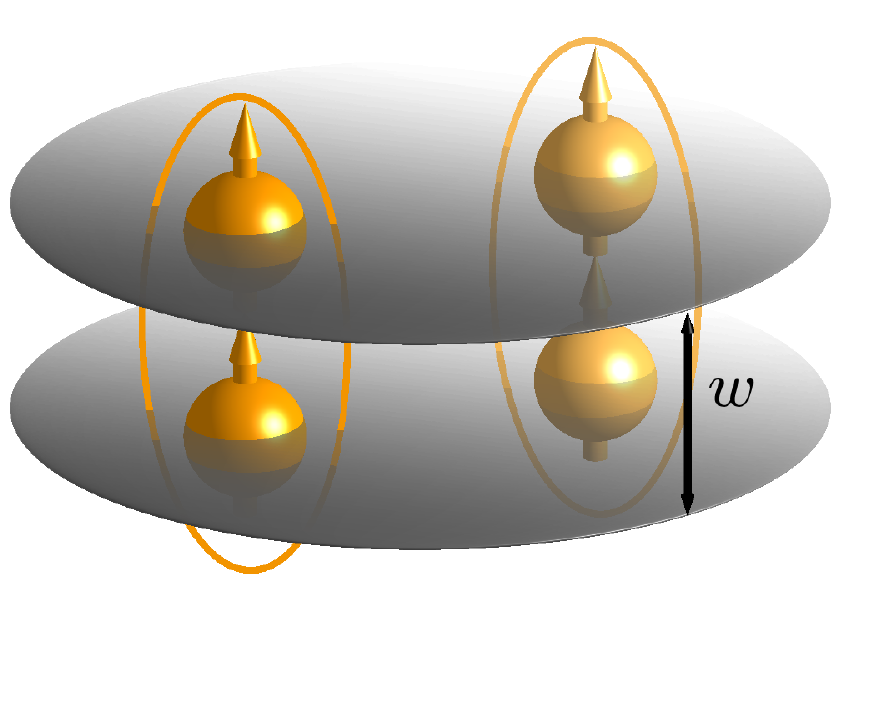}
  \llap{\parbox[b]{8cm}{(a)\\\rule{0ex}{3.cm}}}
  \includegraphics[clip,trim=7.cm 10.cm 19.cm 9.5cm,width=0.48\linewidth]{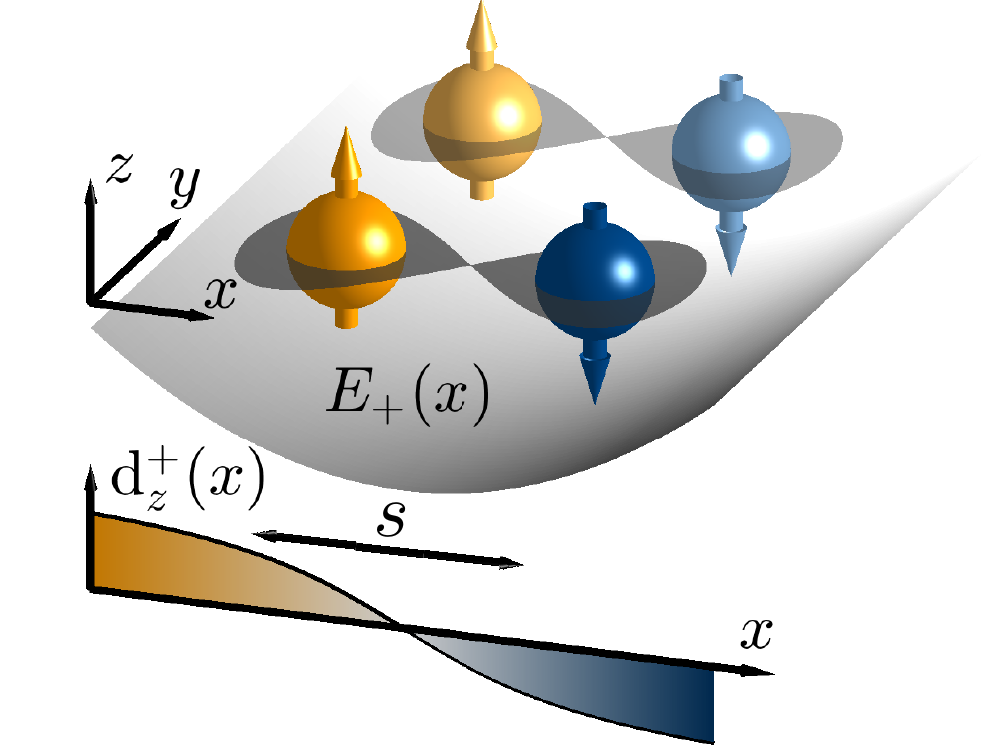}
  \llap{\parbox[b]{7.8cm}{(b)\\\rule{0ex}{3.cm}}}
  \caption{ (a) Polar molecules in a bilayer system in a head-to-tail dipolar
    configuration attract each other, and can form an {\em inter-layer bound
      state}. Here a nanoscale optical barrier separating the two layers
    $w\ll \lambda/2$ provides provides strongly enhanced energy scales for the
    binding energy $E_B$ (see Sec.~\ref{sec:LambdaSystem}).  (b) Strong electric
    field gradients allow the induced electric dipole moment of a molecule $\mathrm{d}_z^+(x)$,
    to change sign on a short spatial distance scale $s$ (see text).  The
    resulting dipolar interaction
    allows the formation of {\em interface bound states} (see
    Sec.~\ref{sec:2LS}). }
  \label{fig:setup}
\end{figure}

Recent experimental progress with ultracold polar molecules~\cite{Park2017,
  Rvachov2017, Reichsollner2017, Seesselberg2018, Petzold2018, Ciamei2018,
  Guo2018, de2019degenerate, Anderegg2019, yang2019singlet, Blackmore2019,
  Yang2019, Voges2019, Tobias2020} opens up unique opportunities to design novel
quantum many-body systems~\cite{Carr2009, Krems2009, baranov2012condensed,
  Bohn2017}. Electric dipole moments, induced by external electric fields in the
manifold of rotational molecular ground states, give rise to long-range and
anisotropic dipolar interactions, which are potentially larger than those
realized with magnetic interactions in atomic systems~\cite{Griesmaier2005,
  Aikawa2012, Lu2012, DePaz2013, Aikawa2014, Naylor2015, Baier2016, Chomaz2018,
  Tang2018, Lepoutre2019, Patscheider2019, Guo2019,Tanzi2019,Tanzi20192}. Thus
polar molecules promise the realization of strongly interacting quantum
many-body systems, e.g., as Hubbard or spin models in optical lattices with
strong nearest-neighbor or long-range interactions~\cite{Micheli2006,
  Capogrosso-Sansone2010, Gorshkov2011, Gorshkov20112}, or in bilayer systems
with strong controllable inter-layer coupling which can be tuned attractive or
repulsive~\cite{Pikovski2010, baranov2011bilayer, Potter2010, Zinner2012}. In an
optical lattice, or a bilayer system created with standing wave laser fields,
the interaction energy between dipolar particles scales as
$E_{{\rm int}}\sim d^{2}/w^{3}$ with $d$ the (induced) dipole moment and $w$ the
lattice- or bilayer spacing provided by $w=\lambda/2$ as half the optical
wavelength~\cite{Pikovski2010, baranov2011bilayer, Potter2010,
  Zinner2012}. However, only polar molecules with the largest electric dipole
moments (LiRb $3.99\,\mathrm{D}$ and LiCs $5.39\,\mathrm{D}$) fulfill the
promise of large off-site interactions approaching the scale of tens of kHz,
comparable or larger than the other relevant energy scales. These interactions
can be quantified by the binding energy $E_B$ of a pair of molecules in a
head-to-tail configuration in a bilayer system formed by a 1D optical lattice
with $w=\lambda/2$ (see Fig.~\ref{fig:setup}(a)).  In addition, $E_B$ must be
larger than available temperatures,
$E_B, E_{{\rm int}}\gtrsim k_{\mathrm{B}}T\approx 1\,\mathrm{kHz}\,
h$~\cite{de2019degenerate}.
For polar molecules with small electric dipole moments, meeting these
requirements can be challenging: For KRb and
$w\equiv \lambda/2 = 250\,\mathrm{nm}$, and assuming $d = 0.33\,\mathrm{D}$ (as
dipole moment in Debye induced by a DC electric field
$12\,\mathrm{kV}/\mathrm{cm}$) implies binding energies less than kHz.
 
Thus, the design of strongly interacting many-body systems with dipole moments
less than a Debye will require, or strongly benefit from going to much smaller
distance scales than those provided by the optical wavelength
scale~\cite{Yi2008,Nascimbene2015,Anderson2019}. % [are these the right refrences; what about our original subwavlength paper with Daley]
In the present paper we explore various possibilities of designing quantum
many-body systems with polar molecules, involving both nanostructured potential
barriers and dipolar interactions modulated on the scale of tens of nanometers,
where the goal is to enhance relevant energy scales. We do this by exploiting
the unique properties offered by polar molecules: This includes the long
life-time of the molecular rotational states in the ground state many-fold, and
the possibilities of controlling induced electric dipole moments of rotational
states.

We will first discuss an all optical scheme (see Sec.~\ref{sec:LambdaSystem}),
were following Refs.~\cite{lkacki2016nanoscale, Budich2017,
  Lacki2019,jendrzejewski2016subwavelength,wang2018dark} for atoms, a nanoscale
barrier can be realized by exploiting the nonlinear response of a molecule
exposed to spatially nonuniform optical light fields in a
$\Lambda$-configuration. These nanostructured barriers allow to split a single
well in a 1D optical lattice, thus forming a bilayer system with separation on a
scale of tens of nanometers.  This leads to significant interactions, and
inter-layer binding energies even for molecules with comparatively small
electric dipole moments.  Assuming $w = 60\,\mathrm{nm}$, the binding energy
becomes $E_B > 10\,\mathrm{kHz}\,h$ for KRb for an induced dipole moment given
above. Table \ref{tab:parametersLambda} provides a list of interaction and
binding energies for various molecules. Besides binding energies, we analyze in
detail loss rates expected for this setup, $\Gamma \approx 10^{-2} E_B$. For
polar molecules there is in addition the opportunity to choose a pair of
rotational ground states in the $\Lambda$-system with induced dipole moments
having opposite sign. This results in {\em repulsive} interactions where one
molecule is {\em inside} the barrier, i.e.,effectively increasing the barrier
height, and a corresponding suppression of the tunneling and thus decay
rate. This can be relevant for chemically reactive molecules~\cite{Krems2009}.

Second, we discuss an all electric scheme (see Sec.~\ref{sec:2LS}), where
instead of the field gradients provided by an optical standing wave we exploit
electric field gradients~\cite{covey2018enhanced}, resulting in
position-dependent energy shifts of molecular states. In contrast to the
nanoscale optical (single-particle) potential barrier discussed above, our aim
is now to modulate the (two-particle) dipolar interaction between molecules on a
short distance scale. The underlying mechanism is to couple a pair of rotational
states with opposite induced dipole moments
$\mathrm{d}_z^{(1)} = -\mathrm{d}_z^{(2)}$ with resonant microwave fields. The
resulting dressed states of this two-level atom acquire position-dependent
dipolar moments
$\mathrm{d}_z^+(x) = \mathrm{d}_z^{(1)}~ \left[(x/s)/\sqrt{1+(x/s)^2}\right]$
illustrated in Fig.~\ref{fig:setup}(b), with the spatial scale of the variation
set by the electric field gradient. Thus we design a two-body interaction
$V(\brho_1,\brho_2) = \mathrm{d}_z^+(x_1) \mathrm{d}_z^+(x_2)/|\brho_1 -
\brho_2|^3$,
where $\rho_{j} = (x_{j}, y_j)$ labels the position of the
$j$\textsuperscript{th} molecule in the $xy$-plane, see Fig.~\ref{fig:setup}(b).
Molecules on different sides of the plane $x=0$ will attract each other allowing
for the formation an {\em interface bound state}. This state is stabilized by
the position dependent dipole-dipole interaction which vanishes the molecules
approach $x=0$, and become repulsive for molecules on the same side. We find
that a bound state occurs above a critical value of dipolar interactions. For
molecules with $d>3\,\mathrm{D}$ (as for LiRb and LiCs) this results in binding
energies of tens of kHz for electric field gradients
$\sim5\,\mathrm{kV}/(\mathrm{cm}\,\mathrm{mm})$. We note that while also in this
scheme achievable binding energies are in principle enhanced by the nanoscale
separation of molecules, this enhancement is partially compensated by the
reduction of the dipolar moment in the vicinity of the interface and the aforementioned scaling of $E_\mathrm{int}$ is
 not applicable in this case. In Table~\ref{tab:parameters} we summarize binding
energies $E_B$ for various molecules for typical parameters. This
all-electric scheme may also be of interest when the optical manipulation of
molecules is not available.

\section{inter-layer bound states in optical \texorpdfstring{$\boldsymbol{\Lambda}$}{Lambda}-systems}
\label{sec:LambdaSystem}

\begin{figure}
\centering
\includegraphics[clip,trim=4.cm 12.5cm 10cm 9.cm,height=.45\linewidth]{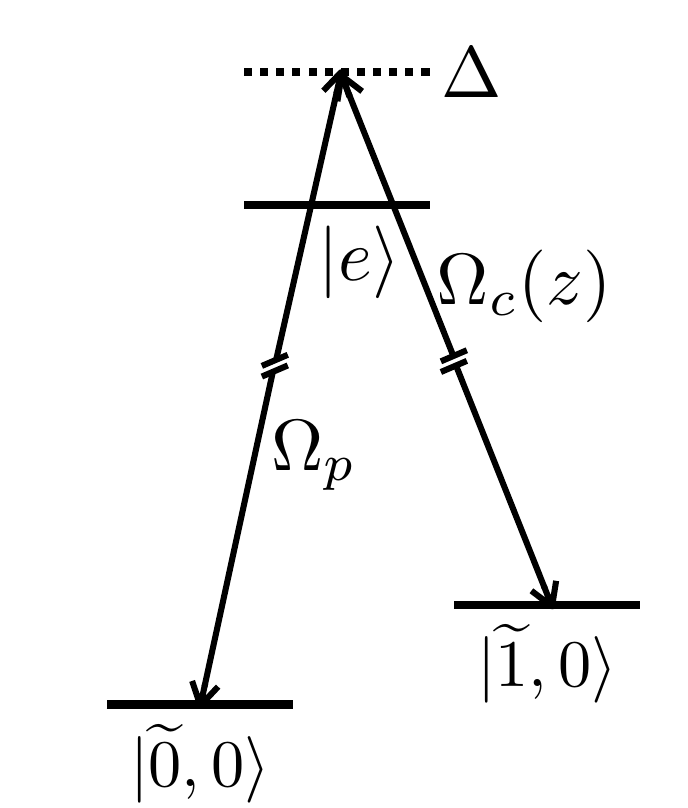}
\llap{\parbox[b]{5cm}{(a)\\\rule{0ex}{3.9cm}}}
\includegraphics[clip,trim=4.cm 8.5cm 4.cm 8.cm,height=.45\linewidth]{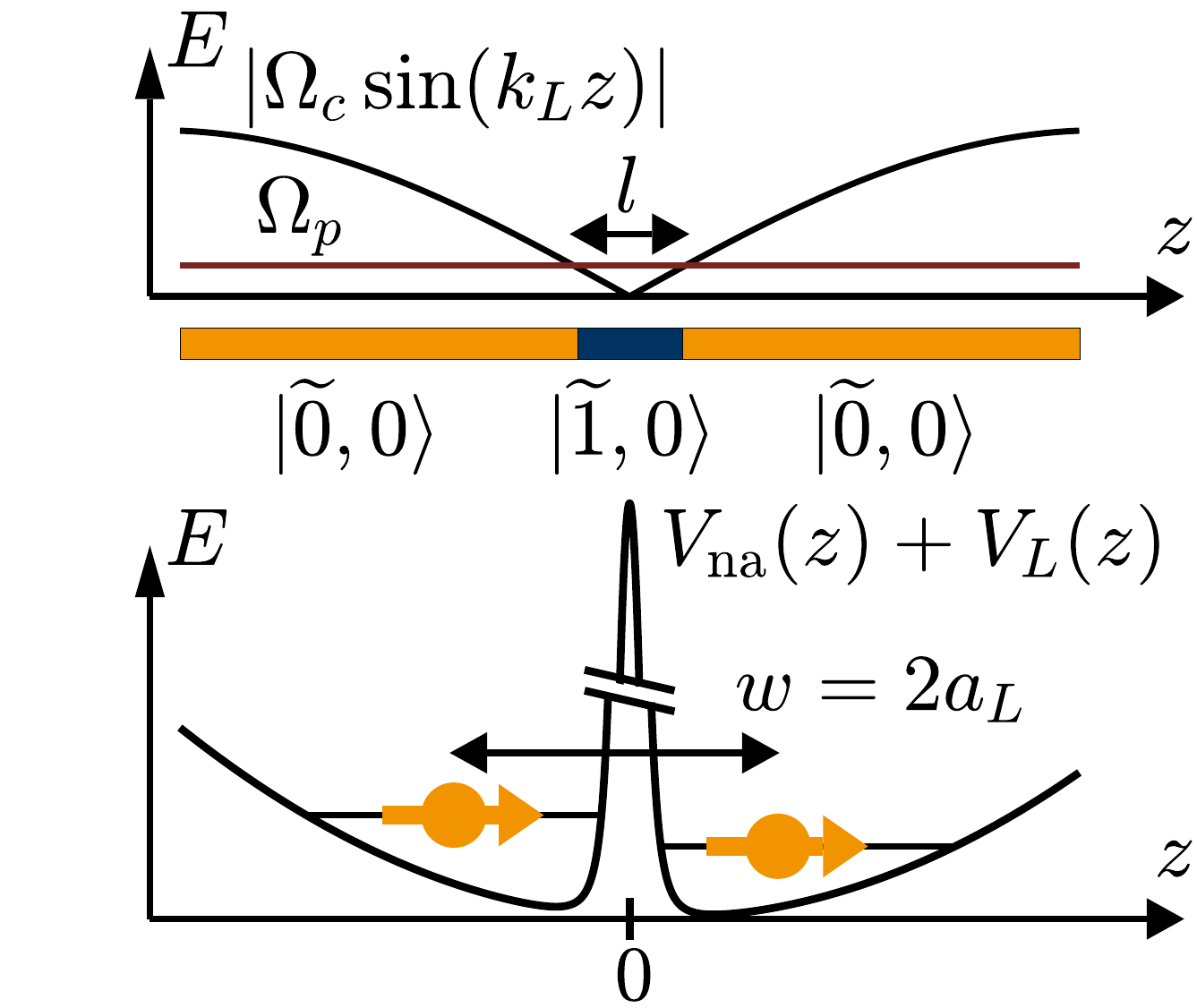}
\llap{\parbox[b]{8.8cm}{(b)\\\rule{0ex}{3.9cm}}}
\llap{\parbox[b]{9cm}{(c)\\\rule{0ex}{1.9cm}}}
\hspace{0.3cm}
\vspace{-0.3cm}
\caption{ (a) Raman coupled $\Lambda$-system. The two rotational states
  $\ket{\widetilde{0},0}$ and $\ket{\widetilde{1},0}$ are coupled by the control
  and probe laser via the electronically excited state $\ket{e}$.  (b) The
  barrier $V_{\mathrm{na}}(z)$ arises as a nonadiabatic correction to the slow
  motion of a molecule in the dark-state Born-Oppenheimer channel of a $\Lambda$-system with
  spatially inhomogeneous Rabi frequencies. The control Rabi beam vanishes at
  $z=0$ which determines the position of the potential barrier
  $V_{\mathrm{na}}(z)$. In the lower panel, the dark-state decomposition as a
  function of position is schematically illustrated. (c) A double well potential
  for dipolar molecules is created by inserting an optical nanoscale barrier
  $V_{\mathrm{na}}(z)$ into a single potential well generated by $V_L(z)$. In
  order to split the potential well into two sites, the width $l$ of $V_L(z)$
  has to be smaller than the ground state size $a_L$ of $V_L(z)$. The effective
  separation of the molecules is $w = 2a_L$.}
  \label{fig:gradient}
\end{figure}

We discuss below bilayer systems for polar molecules with layer separation of
tens of nanometers (see Fig.~\ref{fig:gradient}). Our work builds on earlier
proposals~\cite{lkacki2016nanoscale,jendrzejewski2016subwavelength} and
experiments~\cite{wang2018dark} to realize an optical nanoscale barrier in
$\Lambda$-systems. In Sec.~\ref{sec:Lambda} we first summarize how molecular
$\Lambda$-systems which consist of two rotational levels coupled by a Raman
transition can be trapped in a double-layer geometry with sub-optical-wavelength
spacing as illustrated in Fig.~\ref{fig:gradient}. In Sec.~\ref{sec:Loss} we
give a detailed account of loss channels in such systems, and we identify a
hierarchy of scales which ensures the suppression of losses. Finally, in
Sec.~\ref{sec:Lambda-bound-state} we study the formation of inter-layer bound
states of molecules.

\begin{figure}[t]
\begin{center}
\includegraphics[clip,trim=1.cm 6.5cm 2.2cm 7.cm,height=.35\linewidth]{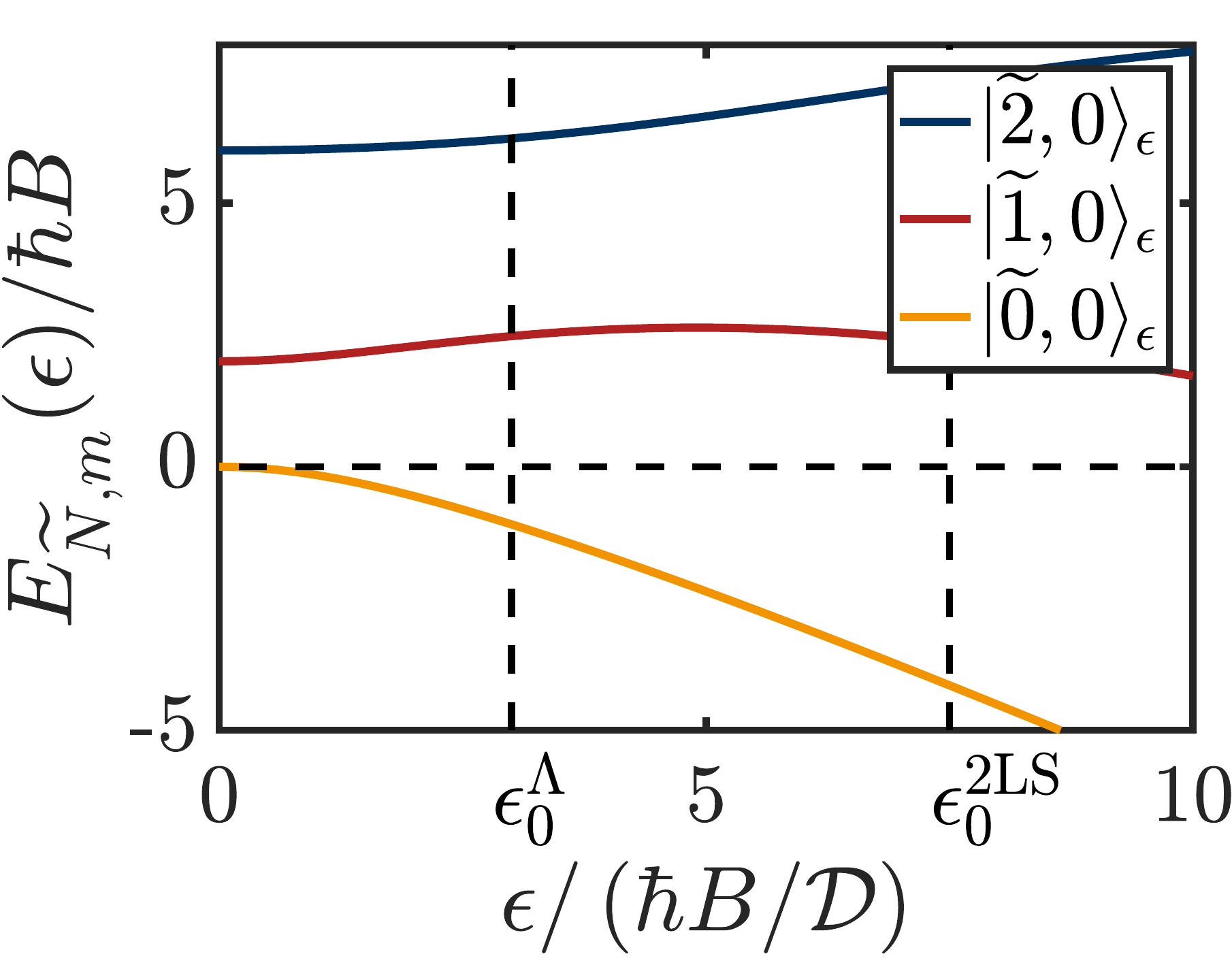}
\llap{\parbox[b]{6.2cm}{(a)\\\rule{0ex}{3.cm}}}
\includegraphics[clip,trim=.8cm 6.5cm 2.2cm 7.cm,height=.35\linewidth]{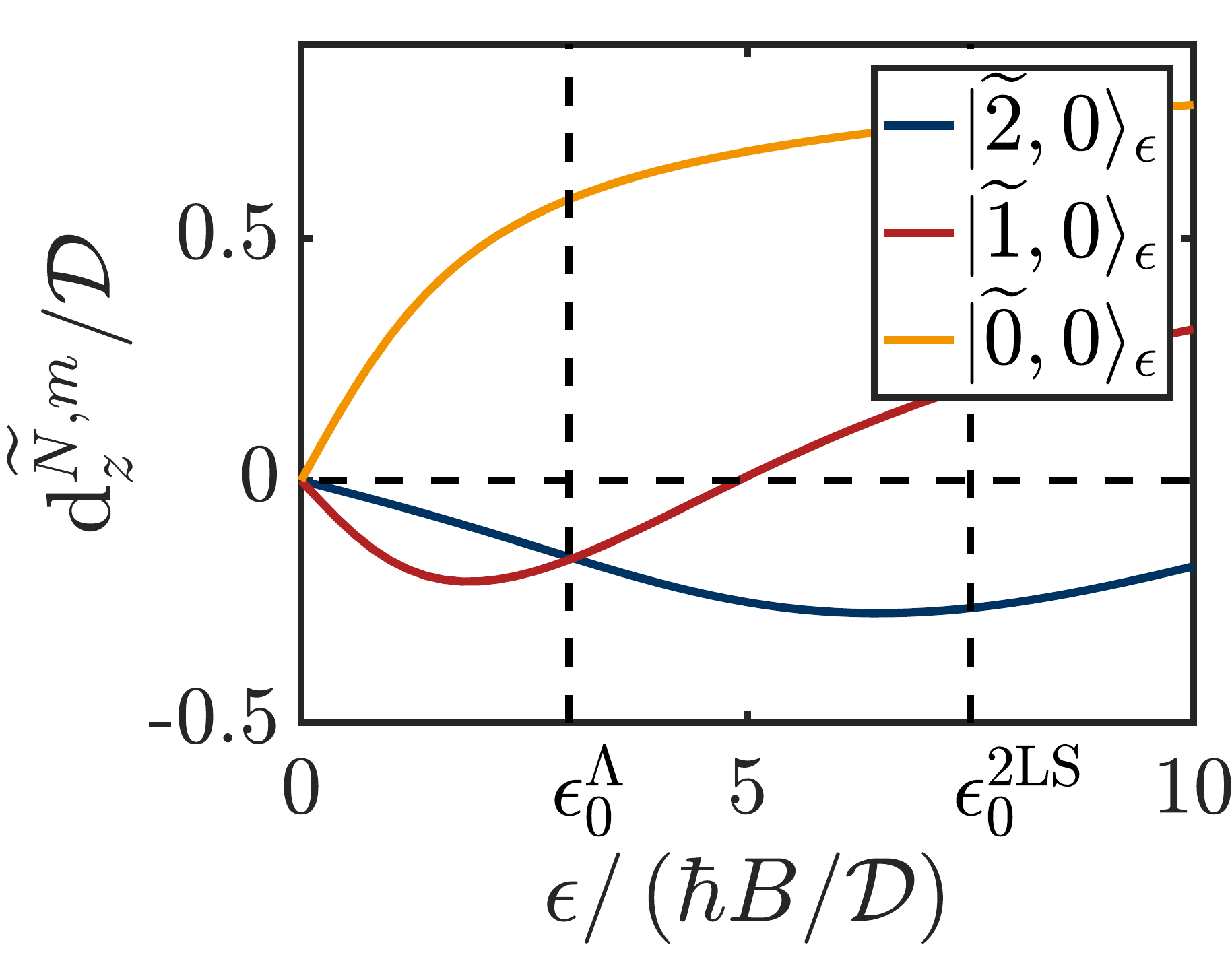}
\llap{\parbox[b]{5.7cm}{(b)\\\rule{0ex}{3cm}}}
\end{center}
\vspace{-0.7cm}
\caption{Energy (a) and induced dipole moment (b) as a function of applied
  electric field for the lowest energy eigenstates of the molecular Hamiltonian
  Eq.~\eqref{eq:Hmol} with $m = 0$. $\epsilon^\Lambda_0$ and
  $\epsilon^{2\mathrm{LS}}_0$ denote the offset fields chosen in
  Secs.~\ref{sec:LambdaSystem} and~\ref{sec:2LS}, respectively.  }
\label{fig:stark}
\end{figure}

The model underlying our discussion is a hetero-nuclear diatomic molecule in its electronic ($X^{1}\Sigma^{+}$ for KRb~\cite{covey2018enhanced}) and vibrational ground state placed in a strong external electric field. Under such conditions, the Stark effect dominates over the hyperfine interactions (this happens already for electric field strengths of tens of V/cm~\cite{Aldegunde2008,Will2016}) and, omitting the latter, the Hamiltonian for a molecule is given by 
\begin{equation}
  H_M = H_R - \mathbf{\hat d }\cdot \boldsymbol{\epsilon},
\label{eq:Hmol}
\end{equation}
where $H_R = \hbar B \hat N^2$ is the Hamiltonian of a rigid rotor, $B$
 the rotational constant and $\hat N$ is the orbital angular momentum
operator. Further, $\be$ is the external electric field, $\mathbf{\hat{d}} = \mathcal{D}\,\hat \br/|\hat\br|$ is the
dipole moment operator expressed in terms of the position operator $\hat \br$,
and $\mathcal{D}$ is the permanent molecule frame dipole moment. 
For $\be = 0$, the eigenstates and energies of $H_M = H_R$ are $\ket{N, m}$ and $E_N = \hbar B N \left( N + 1 \right)$, respectively, where $N = 0, 1, 2, \dotsc$
denotes the angular momentum quantum number, and $m = -N, -N + 1, \dotsc, N$ its
projection onto the quantization axis. Here and in the
following, we choose both the quantization axis and the $z$-coordinate axis
along the direction of the electric field $\be = \epsilon \mathbf{e}_z$, such
that the angular momentum projection quantum number $m$ is conserved. 
For $\be \neq 0$, we denote the eigenstates of $H_M$ by $\ket{\widetilde{N},m}_{\be}$
and the corresponding energy levels $E_{\widetilde N,m}(\be)$ are shown in panel (a) of
Fig.~\ref{fig:stark}. Panel (b) depicts the induced dipole moment which is given
by the change of the energy with the electric field $\be$,
$\mathbf d^{\widetilde{N},m}(\boldsymbol{\epsilon}) = -
\boldsymbol{\nabla}_{\boldsymbol{\epsilon}} E_{\widetilde N,
  m}(\boldsymbol{\epsilon}) = {}_{\be}\bra{\widetilde{N}, m} \hat{\mathbf{d}}
\ket{\widetilde{N}, m}_{\be}$.
The induced dipole moment is aligned along the electric field, i.e., along the
$z$ axis.  From this point on, to simplify the notation, we drop the explicit
dependency on $\be$ for all quantities. 

As mentioned above, for strong electric fields the Stark effect dominates over the hyperfine interactions and, therefore, the nuclear spins become decoupled from the orbital angular momentum. In this case, both $m$ and $m_{1}+m_{2}$, where $m_{1,2}$ is the the nuclear spin projection of the atom $1$ and $2$, respectively, become good quantum numbers~\cite{Aldegunde2008,Will2016}. With an additional magnetic field of a few hundred Gauss, the remaining degeneracy among hyperfine states is lifted, making the individual components $m_{1}$ and $m_{2}$ good quantum numbers. We therefore consider the molecules to be in a single hyperfine state, e.g., the hyperfine groundstate $m_{\mathrm{K}}=-4$ and $m_{\mathrm{Rb}}=3/2$ for \textsuperscript{40}K\textsuperscript{87}Rb~\cite{Ospelkaus2010}.

% \subsection{Lambda-system}
\subsection{Summary of nanoscale potentials in optical \texorpdfstring{$\boldsymbol{\Lambda}$}{Lambda}-systems}
\label{sec:Lambda}

To prepare the ground for the discussion in Secs.~\ref{sec:Loss} and \ref{sec:Lambda-bound-state} and to fix the notation we find it worthwhile for the reader to summarize 
how to engineer a potential barrier on the nanoscale of size $l$ as proposed in the Refs.~\cite{lkacki2016nanoscale} and \cite{jendrzejewski2016subwavelength}. The potential barrier is used to cut a single 
well of an optical potential $V_L(z)$ into two sites as illustrated in Fig.~\ref{fig:gradient}(c). The optical potential can be approximated in the vicinity of its potential minimum $z_0$ by $V_L(z)\approx m \omega_L^2 ( z - z_0 )^2/2$, where $\omega_L$ is the harmonic oscillator frequency. The size of the ground state wave function in this potential is $a_L = \sqrt{\hbar/(m\omega_L)}$ and in the limit $l\ll a_L$ the potential well is split into two sites.
To this end, we consider the motion of a single
molecule along the $z$-axis, which is described by
\begin{equation}
  \label{eq:H-Lambda}
  H_\Lambda = \frac{p_z^2}{2 m} + V_L(z) + H_\Lambda^0(z), 
\end{equation}
where $p_z =-i\hbar\partial_z$ is the $z$-component of the momentum operator
$\mathbf{p} = - i \hbar \nabla$. For the moment, we consider only the motion
along the $z$ axis, and we restore the full 3D form of the Hamiltonian in the
next section. $H_{\Lambda}^0(z)$ is a position dependent $\Lambda$-system
Hamiltonian which is given in a proper rotating frame by
\begin{equation}
H^0_\Lambda = 
\hbar\begin{pmatrix}
-\Delta &\Omega_c(z)/2 & \Omega_p/2 \\
\Omega_c(z)/2 & 0 & 0 \\
\Omega_p/2 & 0 & 0
\end{pmatrix}.
\label{eq:HLambda2}
\end{equation}
As illustrated in Fig.~\ref{fig:gradient}(a), the first leg of the
$\Lambda$-system is a weak control laser $\Omega_p$ which couples the states
$\ket{\widetilde 0, 0}$ and an electronically excited state $\ket{e}$; The
second leg is a strong standing wave control laser
$\Omega_c(z) = \Omega_c\sin(k_cz)$ which couples the states
$\ket{\widetilde 1, 0}$ and $\ket{e}$. $\Delta$ denotes the detuning of the
Raman transition, see Fig.~\ref{fig:gradient}(a) and (b). In the case of KRb, a
possible candidate for $\ket{e}$ is a vibrational excitation of the $1^1\Pi$
electronically excited state~\cite{Wang2010}, for which the Franck-Condon factor
is maximized~\cite{Okada1996}. Rabi coupling between the state $1^1\Pi$ and the
absolute ground state of KRb has been demonstrated in Ref.~\cite{Wang2010} in
the absence of an electric offset field. We discuss the decay of molecules which
results from the finite lifetime of the electronically excited state in
Sec.~\ref{sec:Loss}.

A Raman coupled $\Lambda$-system always hosts one dark state at zero energy and
two bright states, which we denote by $\ket{0}_z$ and $\ket{\pm}_z$,
respectively. The corresponding eigenenergies are given by
\begin{equation}
\label{eq:brightEnergy}
E^\Lambda_0 = 0\mathrm{ ~and~ } E^\Lambda_\pm = \frac{\hbar}{2}\left[-\Delta \pm \sqrt{\Omega_p^2 +\Omega_c^2(z) + \Delta^2} \right]. 
\end{equation}
and in particular, the dark state reads
\begin{equation}
  \label{eq:zero-energy-BO-state}  
  \ket{0}_z = \frac{1}{\sqrt{1+(z/l)^2}}\left(z/l \ket{\widetilde 0, 0} -  \ket{\widetilde 1, 0}\right),
\end{equation}
where we used that $k_c|z-z_0|\lesssim k_c a_L\ll 1$. The characteristic length scale on which the structure of $\ket{0}_z$
changes is given by
\begin{equation}
\label{eq:charlength}
l = \Omega_p/(\Omega_c \ k_c).
\end{equation}
For $\lvert z \rvert \gg l$, the dark state is essentially equal to the internal
state $\ket{\widetilde 0, 0}$; The contribution to $\ket{0}_z$ from
$\ket{\widetilde{1}, 0}$ is relevant only in a region of size $l$ around the
zero crossing of $\Omega_c(z)$ as illustrated in Fig.~\ref{fig:gradient}(b).

We consider the limit of slow motion of the molecules, in which the eigenvalues
and the eigenstates of $H_\Lambda^0(z)$ form decoupled Born-Oppenheimer (BO)
channels~\cite{burrows2017nonadiabatic, kazantsev1990mechanical, dum1996gauge,
  dutta1999tunneling, ruseckas2005non}. Nonadiabatic corrections yield two 
  types of contributions: On the one hand, they describes nonadiabatic
channel couplings; on the other hand, they gives rise to repulsive potential
barriers. Below, we discuss conditions under which the nonadiabatic channel
couplings are negligible. Then, the Hamiltonian for the motion of a molecule
prepared in the dark state, i.e., the zero-energy channel, can be written as
  \begin{equation}
  \label{eq:zero-energy-BO-channel}
  H_{\mathrm{\Lambda}, 00}^{\mathrm{ad}} = \frac{p_z^2}{2 m} + V_L(z) +
  V_{\mathrm{na}}(z),
\end{equation}
where the nonadiabatic potential barrier is independent on the value of $\Delta$ and reads 
\begin{equation}
  \label{eq:Lambda-na-potential-barrier}
  V_{\mathrm{na}}(z) =
  \frac{\hbar^2}{2m l^2} \frac{1}{[1+(z/l)^2]^2},
\end{equation}
as introduced in Refs.~\cite{lkacki2016nanoscale} and \cite{jendrzejewski2016subwavelength}, see also Appendix \ref{sec:basistrans}. The characteristic scale $l$ from Eq.~\eqref{eq:charlength}, which defines both
the width and the height of the potential barrier $V_{\mathrm{na}}(z)$, is
determined by the Rabi frequencies $\Omega_p$ and $\Omega_c$ and the wave vector
$k_c$, see Fig.~\ref{fig:gradient}. By choosing these parameters such that
$a_L\gg l$, a single potential well generated by $V_L(z)$ is split into two
sites. This is illustrated in Fig.~\ref{fig:leak1}(a), where we show the ground
($\psi_R(z)$) and first excited ($\psi_L(z)$) state in the zero-energy BO
channel, which are located, respectively, to the right and left of the barrier.
We obtain the states $\psi_{L, R}(z)$ by numerically diagonalizing the
Hamiltonian in Eq.~\eqref{eq:zero-energy-BO-channel}. An analytical discussion
of the suppression of wave functions inside the barrier is provided in
Appendix~\ref{sec:decayBO}.

\subsection{Loss channels and hierarchy of scales}
\label{sec:Loss}
The Hamiltonian Eq.~\eqref{eq:zero-energy-BO-channel} ignores nonadiabatic
channel couplings. These couplings induce decay of states in the zero-energy BO
channel to the other channels. In the vicinity of $z = 0$, the energies of the
bright states give rise to trapping and antitrapping of molecules in the $+$ and
$-$ channels, respectively. Therefore, at low energies, the $+$ channel hosts a
discrete set of trapped states. None of these states are resonant with the
states $\psi_{L, R}(z)$ in the zero-energy channel if the minimal gap
$\hbar \Omega_p/ 2$ between the zero and the $+$ channels is larger than the
spacing of levels $\approx \hbar \omega_L$ in the zero-energy channel. In
contrast, the antitrapped states in the $-$ channel form a continuum, and
resonant transitions between the states $\psi_{L,R}(z)$ and states in the $-$
channel are possible. The decay rate from the zero-energy BO channel to the $-$
channel can be obtained by using Fermi's golden rule. The derivation presented
in Appendix~\ref{sec:decayBO} yields for $\Delta=0$
\begin{equation}
 \label{eq:Lambda-decay-rate}
 \frac{\Gamma_{\Lambda}}{\omega_{L}}=2.5\beta^{2}\frac{l}{a_{L}}\sqrt{\kappa}~\mathrm{exp}\left(-1.75~\kappa\right),
\end{equation}
where $\kappa$ is the square root of the ratio of the gap between the
zero-energy and $-$ channels and the height of the barrier,
$\kappa=(\hbar\Omega_{p}/2)/(\hbar^{2}/2ml^{2})$, and the number $\beta\leq0.34$
is determined by the exact position of the barrier and the wave function of the
trapped state. This result is valid in the limit $\kappa\gg1$ or equivalently
$l/a_{L}\gg\sqrt{\omega_{L}/\Omega_{c}}$, when $\Gamma_{\Lambda}$ is
exponentially suppressed. The gap between the zero-energy and $-$ channels and
thus the parameter $\kappa$ is enhanced in the far blue-detuned regime in which
$\Delta \gg \lvert \Omega_{p, c} \rvert$, and where according to
Eq.~\eqref{eq:brightEnergy} the size of the gap is equal to $\hbar \Delta$. 
In this case, Eq.~\eqref{eq:Lambda-decay-rate} provides only an estimate of the decay rate, 
because it does not take into account the change of the non-adiabtic couplings with $\Delta$.

In contrast to rotational excitations, electronically excited states have a
non-negligible decay rate $\gamma_e$. If the motion of molecules in the
zero-energy or dark-state channel is perfectly adiabatic, they are not affected
by this decay. However, nonadiabatic corrections couple the dark-state channel
to the $\pm$ or bright-state channels, and thus lead to a small admixture of the
electronically excited state $\ket{e}$ to the dark state, which in turn can lead
to inelastic scattering of photons. This
effect is strongly suppressed in the far blue-detuned regime: On the one hand,
for $\Delta \gg \lvert \Omega_{p, c} \rvert$, the $-$ BO-channel is shifted down
energetically, and the mixing between the dark-state and $-$ channels is
negligible. On the other hand, the gap between the $+$ and zero-energy channels
is reduced. To estimate the mixing between the dark-state and $+$ channels, we
first analyze wave functions in the $+$ channel. For
$\Delta \gtrsim |\Omega_{p,c}|$ the $+$ bright-state energy in
Eq.~\eqref{eq:brightEnergy} can be approximated around $z = 0$ as a harmonic
potential, $E_+(z)\approx \Omega_c^2/(4\Delta) + m\omega_+^2 z^2/2$, where
$\omega_+ = \Omega_c \sqrt{\hbar^2k_c^2/(2m\Delta)}$ takes the role of the
harmonic oscillator frequency. The $+$ channel hosts a barrier which is for far
blue detuning equivalent to the one in the zero-energy channel. Therefore, as
discussed in Appendix~\ref{sec:diag-nonad-corr}, eigenfunctions in the $+$
channel are suppressed within the barrier by a factor of $l/a_+$, where
$a_+ = \sqrt{\hbar/(m\omega_+)}$ is the oscillator length. The admixture of $+$
channel wave functions can be approximated to first order in the diabatic
corrections by $c_+\approx (l/a_L) (l/a_+) V_\mathrm{na}(0)/\Delta E$, where the
factors $l/a_L$ and $l/a_+$ describe the reduction of the dark-state and
bright-state wave functions, respectively. The coupling matrix element is
approximately given by the height of the barrier
$V_\mathrm{na}(0)$~\cite{jendrzejewski2016subwavelength}, and $\Delta E$ is the
energy gap. For $\Delta > \Omega_c$ the gap o the lowest energy state in the $+$ BO-channel 
is $\Delta E\approx \hbar\omega_+/2$, which yields a relatively small admixture,
$|c_+|^2< 0.1$ for $l/a_L = 0.1$. In the far blue-detuned limit, the amplitude of the excited
state $\ket{e}$ in the $+$ channel is small and therefore a reduced decay rate
 $\gamma_e^+ \approx \gamma_e \Omega_p^2/(8\Delta^2)$ (see Sec. B2 in the
supplementary material of Ref.~\cite{lkacki2016nanoscale}) has to be used to
estimate the inelastic scattering rate in the zero-energy BO channel,
$\gamma_e^0\approx \gamma_e^+ |c_+|^2 \approx 10^{-4} \gamma_e$, where $\gamma_e$
is typically on the order of tens of MHz~\cite{Wang2010, Ni2008}. Inelastic
light scattering is negligible when $\gamma_e^0\ll\omega_L$, which is indeed the
case for $\omega_L \approx2\pi\times 88\,\mathrm{kHz}$ as considered in Sec.~\ref{sec:LambdaParameters}.

\begin{figure}
\centering
\includegraphics[clip,trim=1.cm 6.5cm 0.cm 7.cm,height=.43\linewidth]{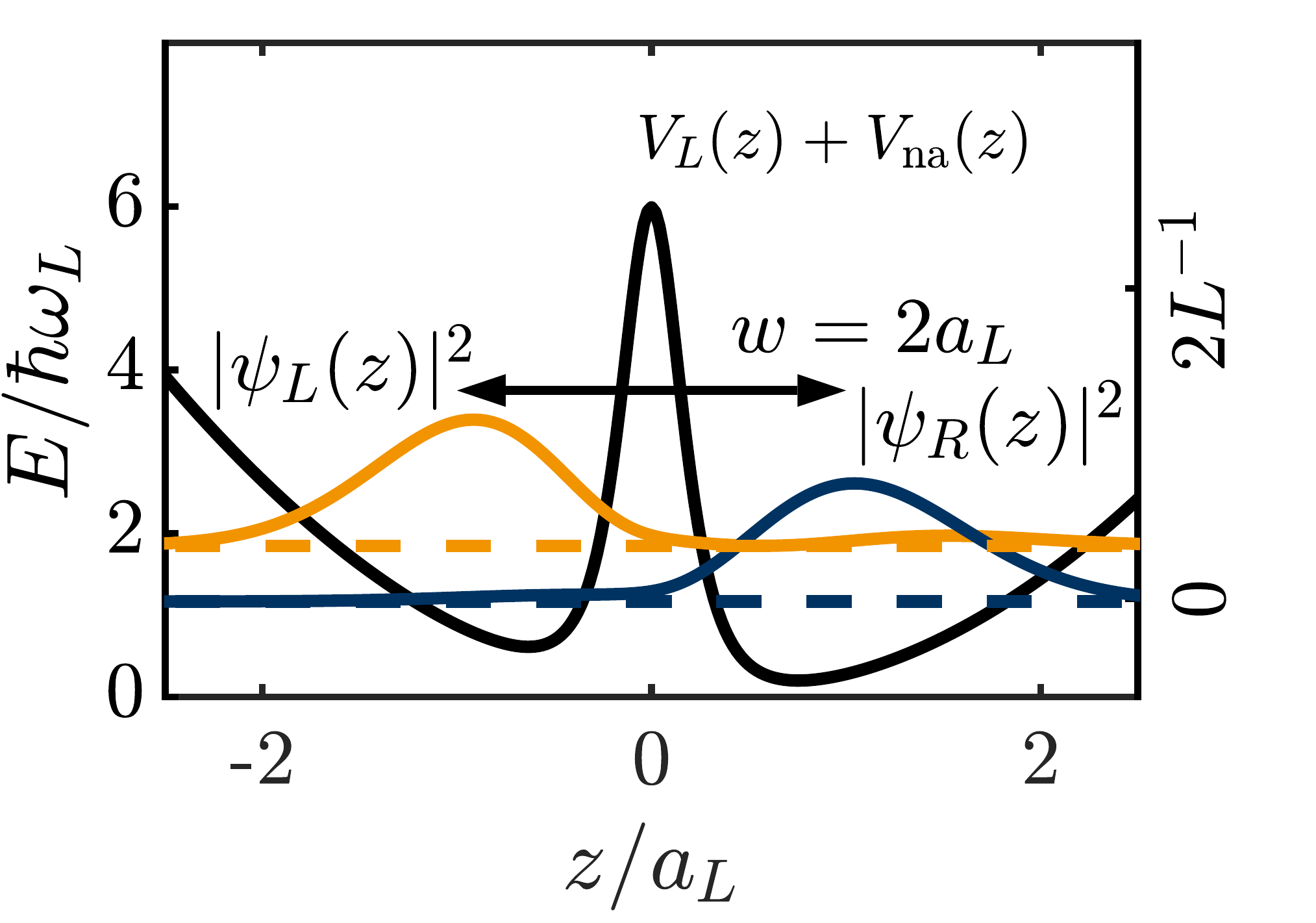}
\llap{\parbox[b]{9.1cm}{(a)\\\rule{0ex}{3.6cm}}}
\hspace{-0.17cm}
\includegraphics[clip,trim=5.cm 6.5cm 4.6cm 7.cm,height=.43\linewidth]{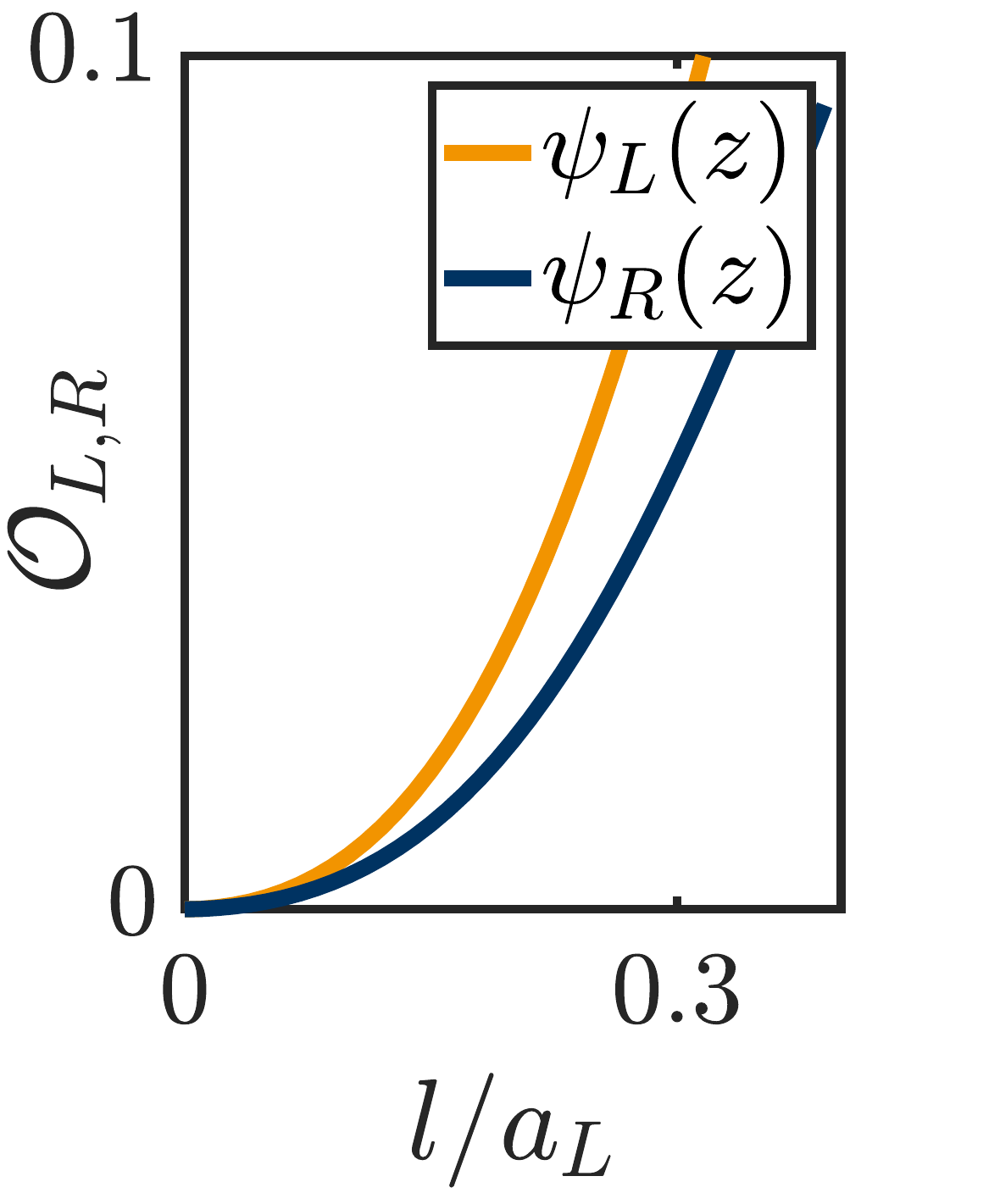}
\llap{\parbox[b]{4.8cm}{(b)\\\rule{0ex}{3.6cm}}}
\vspace{-0.4cm}
\caption{(a) Spatial probability densities for the ground
  ($\psi_R(z)$) and first excited state ($\psi_L(z)$) in the zero-energy BO
  channel. The dashed baselines indicate the corresponding eigenenergies. The
  peaks of the wave functions are separated by $w = 2 a_L$.
  Parameters are $l/a_L = 0.3$ and $z_0/a_L = 0.3$. Energies and the probability
  density of $\psi_R(z)$ correspond to the left-hand and right-hand axis,
  respectively, and $L$ denotes the numerical box size. The scale for the
  probability density of $\psi_L(z)$ is the same as for $\psi_R(z)$, shifted to
  the corresponding baseline. (b) Leakage for $\psi_L(z)$ and $\psi_R(z)$
  through the barrier as a function of $l/a_L$ with fixed ratio
  $z_0/a_L = 0.3$.}
\label{fig:leak1}
\end{figure}

In addition to the decay of the states $\psi_{L, R}(z)$ in the zero-energy BO
channel due to non-adiabatic channel couplings and inelastic light scattering, the
stability of a many-body system that is loaded into the states $\psi_{L, R}(z)$
is reduced by inelastic collisions, like chemical reactions \cite{Micheli2010}, which can occur at short distances. This
effects are minimized when $\psi_L(z)$ and $\psi_R(z)$ are maximally localized on
either side of the barrier. As a measure of the degree of localization we
calculate the wave function leakage
$\mathcal O_{L,R} = \int_{0,-\infty}^{\infty,0} \text{d} z \, \lvert
\psi_{L,R}(z) \rvert^2$.
The leakage is affected by the length scales $a_L$, $l$ and $z_0$, which
determine the shape of the wave functions. For $z_0/a_L = 0.3$, in
Fig.~\ref{fig:leak1}(c) we observe minimal leakage for a range of small values
of $l/a_L$. In particular, leakage is negligible in the limit $l/a_L \ll 1$. 

Even when leakage of the eigenstates of the single-particle
Hamiltonian~\eqref{eq:zero-energy-BO-channel} is strongly suppressed, molecules
which are loaded into the potential well on one side of the barrier can tunnel
through the barrier and collide inelastically with molecules on the other
side. Tunneling through the barrier can be estimated by the transmission
coefficient $t \approx \sqrt{4E/(\pi^2V_\mathrm{na}(0))}$ for an incoming plane
wave with energy $E \ll V_\mathrm{na}(0)$~\cite{Lacki2019}. The tunneling rate
is then given by $J = t^2E/h$, where $t^2$ is the probability for a molecule to
tunnel through the barrier and $E/h$ is the attempt frequency. For a particle
prepared in $\psi_{L,R}(z)$, we set $E = E_{L,R}$, which yields a tunneling
constant of $J_{L,R}/\omega_L = 8/\pi^3(l/a_L)^2
E_{L,R}^2/(\hbar\omega_L)^2$.
For a stable bilayer system we require $J_{L,R}/\omega_L\ll 1$ which is
achievable for $l/a_L \ll 1$. In particular, for $l/a_L = 0.1$ we get
$J_{L,R}/\omega_L\leq 0.01$.

We obtain a hierarchy of conditions which have to be met by combining the
requirements of small leakage, tunneling and decay
rates~\eqref{eq:Lambda-decay-rate}:
\begin{equation}
  1\gg \frac{l}{a_L}\gg\sqrt{\frac{\omega_L}{\Omega_c}}.
  \label{eq:inequality}
\end{equation}
Engineering barriers to cut a single well into two sites is possible if the
above states hierarchy of scales is fulfilled. We emphasize that our scheme has
sufficiently many ``tuning knobs'' to adjust each parameter independently. In
the next section, we discuss many-body effects arising from dipolar interactions
across the barrier.

\subsection{Dipolar interaction and inter-layer bound state}
\label{sec:Lambda-bound-state}

We now turn to many-body physics of molecules in the $\Lambda$ configuration
discussed above. As an illustrative example, we consider the motion of two
identical fermionic molecules. In particular, we study the formation of bound
states between molecules on the left and right of the optical nanoscale
barrier. We assume a 3D geometry, in which the motion of molecules is not
restricted in the $xy$ plane. For molecules in the zero-energy BO channel, the
position-dependent internal state is given by $\ket{0}_z$ in
Eq.~\eqref{eq:zero-energy-BO-state}. The induced dipole moment of this state is
oriented along the static electric field, i.e., along the $z$ axis. Therefore,
the interaction of molecules within one of the layers $L$ or $R$ is always
repulsive~\cite{note:hyperfine}, whereas molecules in different layers experience attractive
interactions when their separation
$\boldsymbol{\rho} = x \mathbf{e}_x + y \mathbf{e}_y$ in the $xy$ plane has
sufficiently small magnitude $\rho = \sqrt{x^2 + y^2}$, so that their relative
position corresponds to a head-to-tail configuration. As illustrated in
Fig.~\ref{fig:setup}, this can lead to the formation of a bound state.

To describe the bound state quantitatively, we start with the Hamiltonian for
the motion of two molecules $j = 1,2$ with position coordinates
$\br_j = (x_j, y_j, z_j)$ in the 3LS configuration described in the previous
section, which reads
\begin{align}
  \label{eq:H-two-body-Lambda}
  &H_{2, \Lambda} = \sum_{j=1,2} \left( \frac{\mathbf{p}_j^2}{2m} + V_L(z_j) +
    H^0_{\Lambda}(z_j) \right) \\ 
    &+ U^{(1)}_{\Lambda}(t)\otimes
  U^{(2)}_{\Lambda}(t) ~ V_{\mathrm{dd}}(\br_1,\br_2) ~
  U_{\Lambda}^{(2),\dagger}(t) \otimes U_{\Lambda}^{(1),\dagger}(t),\notag
\end{align}
where the dipolar interaction is given by 
\begin{equation}
  \label{eq:V-dd}
  V_{\mathrm{dd}}(\br) = \frac{1}{|\br|^3} \left[\hat\bd_1
    \cdot\hat\bd_2 - 3 \frac{\hat\bd_1 \cdot \br
      ~\hat\bd_2 \cdot \br }{|\br|^2} \right]
\end{equation}
and
$U^{(j)}_{\Lambda}(t) = \exp\left[-i\omega_1 t \ket{\widetilde
    0,0}\bra{\widetilde 0,0} -i\omega_2 t \ket{\widetilde 1,0}\bra{\widetilde
    1,0} \right]$
is the rotating-frame transformation acting on the $j$\textsuperscript{th}
particle. The frequencies $\omega_{1, 2}$ are chosen such that
$\Delta = \omega_1 - (E_e - E_{\widetilde 0,0})/\hbar  = \omega_2-(E_e - E_{\widetilde 1,0})/\hbar$
where $E_e$ denotes the energy of the electronically excited state.

When we project the Hamiltonian in Eq.~\eqref{eq:H-two-body-Lambda} to the
zero-energy channel, the contribution $H_{\Lambda}^0(z_j)$ is replaced by the
effective BO potential in Eq.~\eqref{eq:Lambda-na-potential-barrier}. The
dipolar interaction for two molecules in the zero-energy BO channel reads
\begin{equation}
\begin{split}
  V_{\mathrm{dd}}^0(\mathbf{r}_1, \mathbf{r}_2) =& {}_{z_1}\bra{0}
  \otimes{}_{z_2}\bra{0} \left[ U^{(1)}_{\Lambda}(t)\otimes U^{(2)}_{\Lambda}(t)\right. \\ 
   \times V_{\mathrm{dd}}(\br_1,\br_2)& ~\left.
     U_{\Lambda}^{(2),\dagger}(t)\otimes U_{\Lambda}^{(1),\dagger}(t) \right]
  \ket{0}_{z_1} \otimes\ket{0}_{z_2}\\
  \approx&  \frac{\mathrm{d}_z(z_1)\mathrm{d}_z(z_2)}{r_{1,2}^3}
  \left[1-3\frac{(z_1-z_2)^2}{r_{1,2}^2} \right],
  \end{split}
\end{equation}
where $r_{1,2}^2 = (z_1 - z_2)^2 + \rho^2$ denotes the relative distance of the
two molecules in 3D, and
\begin{equation}
  \hspace{-0.2cm}
  \mathrm{d}_z(z) = {}_z\bra{0}\mathrm{d}_z\ket{0}_z = \frac{1}{1+(z/l)^2} \left[ (z/l)^2 \mathrm{d}^{\widetilde 0, 0}_z + \mathrm{d}^{\widetilde1,0}_z\right].
\end{equation} 
The above expression for the dipolar interaction is valid if fast oscillating
terms are neglected and the penetration of the single-particle wave functions
into the barrier is negligible which is the case for $l/a_L \ll 1$. In summary,
the projection of the two-molecule Hamiltonian~\eqref{eq:H-two-body-Lambda} to
the zero-energy channel reads
\begin{equation}
  \label{eq:H-two-body-Lambda-0-channel}
  \hspace{-0.22cm}
  H_{2, \Lambda}^0 = \sum_{j=1,2} \left[ \frac{\mathbf{p}_j^2}{2m} +
    V_L(z_j) + V_{\mathrm{na}}(z_j) \right] +
  V_{\mathrm{dd}}^0(\br_1,\br_2).
\end{equation}
We assume in the following that the single-particle energy scales associated
with the nanoscale potential are dominant as compared to the dipolar
interaction, such that motion of the two molecules in the regime of low energies
is restricted to the states $\psi_L(z)$ and $\psi_R(z)$. Under these conditions,
the nanoscale potential effectively implements a two-layer geometry, with layers
$L$ and $R$ which are parallel to the $xy$ plane. As illustrated in
Fig.~\ref{fig:leak1}(a), the separation of the peaks of the wave functions
defines an effective layer separation given by $w \approx 2 a_L$.
The two-body wave function $\Psi(\mathbf{r}_1, \mathbf{r}_2)$ in the zero-energy
channel can then be written as
\begin{equation}
  \label{eq:Psi-Lambda}
  \Psi(\mathbf{r}_1, \mathbf{r}_2) = \Psi_{\parallel}(\boldsymbol{\rho}_1,
  \boldsymbol{\rho}_2) \sum_{\alpha, \beta = L, R} c_{\alpha \beta}
  \, \psi_{\alpha}(z_1) \, \psi_{\beta}(z_2).
\end{equation}
The component $\Psi_{\parallel}(\boldsymbol{\rho}_1, \boldsymbol{\rho}_2)$,
which describes motion in the $xy$ plane, can be decomposed further: First, the
system is translationally invariant, and therefore
$\Psi_{\parallel}(\boldsymbol{\rho}_1, \boldsymbol{\rho}_2) =
\Psi_{\mathrm{rel}}(\boldsymbol{\rho}) \Psi_{\mathrm{COM}}(\mathbf{R})$
factors into contributions corresponding to the relative and COM motion with
respective coordinates
$\boldsymbol{\rho} = \boldsymbol{\rho}_1 - \boldsymbol{\rho}_2$ and
$\mathbf{R} = (\boldsymbol{\rho}_1 + \boldsymbol{\rho}_2)/2$. Second, due to the
symmetry of the Hamiltonian under rotations around the $z$ axis, the wave
function corresponding to the relative motion can be decomposed into radial and
angular components,
$\Psi_{\mathrm{rel}}(\boldsymbol{\rho}) = \sum_{m_z \in \Z} \chi_{m_z}(\rho)
e^{i m_z \phi}$.
To find a bound state it is sufficient to consider $m_z = 0$, since a
nonvanishing angular momentum adds an additional repulsive centrifugal barrier,
which increases the energy. Then, antisymmetry of the wave function
$\Psi(\mathbf{r}_1, \mathbf{r}_2)$ under the exchange of particles requires that
the motional state of the two molecules along the $z$ axis is an antisymmetric
superposition of the single-particle states $\psi_L(z)$ and $\psi_R(z)$, that
is, the coefficients $c_{\alpha \beta}$ in Eq.~\eqref{eq:Psi-Lambda} are given
by $c_{LL} = c_{RR} = 0$ and $c_{LR} = - c_{RL} = 1/\sqrt{2}$, so that
\begin{multline}
  \Psi(\mathbf{r}_1, \mathbf{r}_2) = \frac{1}{\sqrt{2}} \chi_0(\rho)
  \Psi_{\mathrm{COM}}(\mathbf{R}) \\ \times \left[ \psi_L(z_1) \psi_R(z_2) -
    \psi_R(z_1) \psi_L(z_2) \right].
\end{multline}
With this ansatz, the Hamiltonian~\eqref{eq:H-two-body-Lambda} yields the
following Schr\"odinger equation (SE) for the radial component of the two-body
wave function:
\begin{equation}
  \left[ - \frac{\hbar^2}{2 m} \left( \frac{\partial^2}{\partial \rho^2} + \frac{1}{\rho}
      \frac{\partial}{\partial \rho} \right) + V_\mathrm{2D}(\rho)
  \right] \chi_0(\rho) = E_B \chi_0(\rho).
  \label{eq:SEEB}
\end{equation}
The effective 2D dipolar interaction is given by
\begin{multline}
  \label{eq:V-2D}
  V_{\mathrm{2D}}(\rho) = \frac{1}{2} \int \mathrm{d} z_1 \mathrm{d} z_2 ~
  V_{\mathrm{dd}}^0(\mathbf{r}_1, \mathbf{r}_2) \\ \times \left[ \psi_L(z_1)
    \psi_R(z_2) - \psi_R(z_1) \psi_L(z_2) \right]^2.
\end{multline}
It can be decomposed into ``direct'' and ``exchange'' contributions,
$V_{\mathrm{2D}}(\rho) = V_\mathrm{2D}^{\mathrm{D}}(\rho) -
V_\mathrm{2D}^{\mathrm{E}}(\rho)$, which read
\begin{multline}
  V^{\mathrm{D,E}}_\mathrm{2D}(\rho) = \int\mathrm{d}z_1\mathrm{d}z_2 \,
  \psi_{L,L}(z_1)\psi_{R,R}(z_2) \\ \times V_{\mathrm{dd}}(\rho, z_1, z_2)
  ~\psi_{R,L}(z_2)\psi_{L,R}(z_1),
\end{multline}
In the limit of small spatial wave function overlap which is realized for
$l/a_L\ll 1$, the exchange part of the interaction gives a strongly suppressed
repulsive contribution which we neglect in the following.

\begin{figure}[t]
\centering
\includegraphics[clip,trim=3.7cm 6.6cm 5.5cm 6cm,height=.47\linewidth]{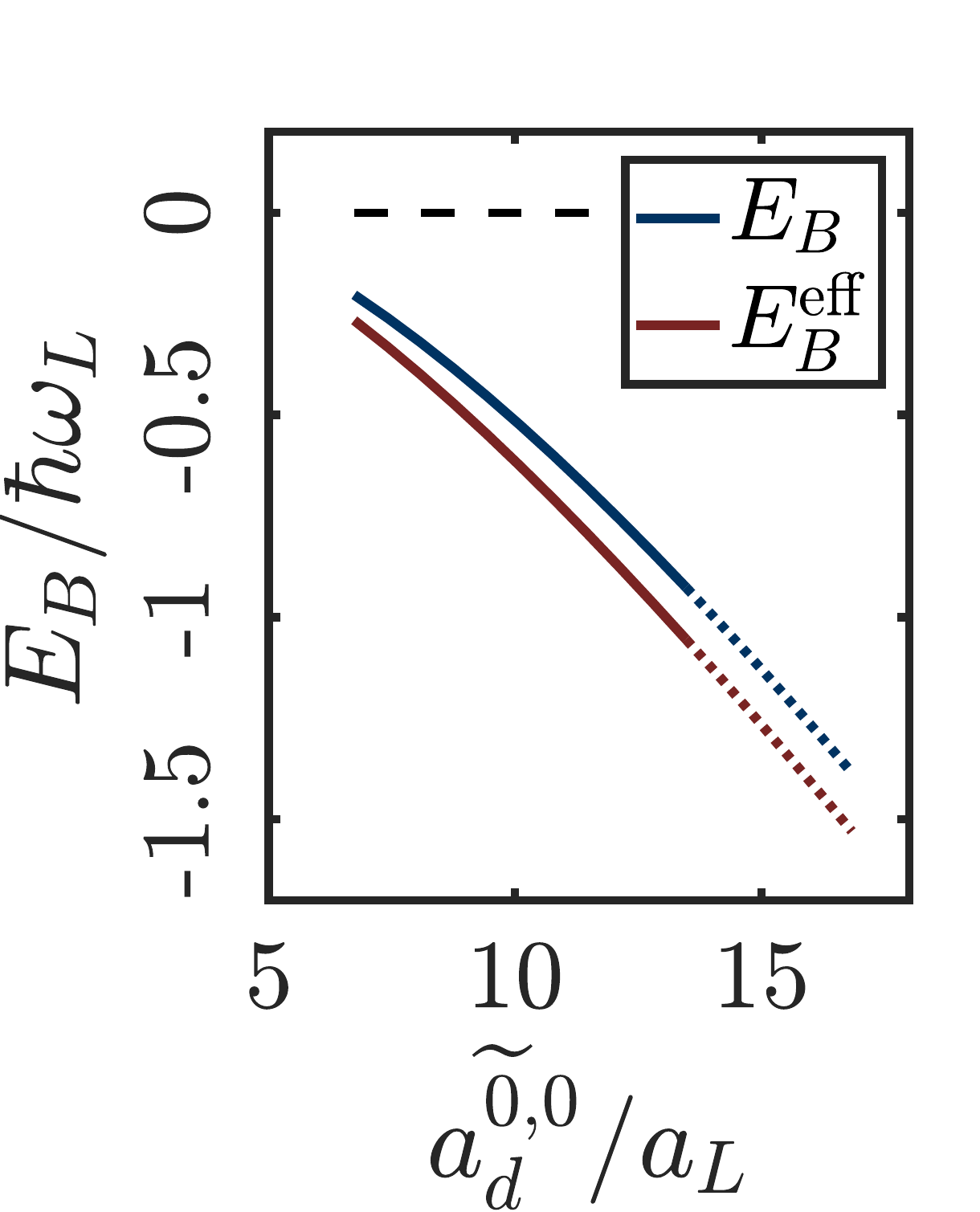}
\llap{\parbox[b]{4.6cm}{(a)\\\rule{0ex}{3.8cm}}}
\includegraphics[clip,trim=1cm 6.cm 1cm 6cm,height=.47\linewidth]{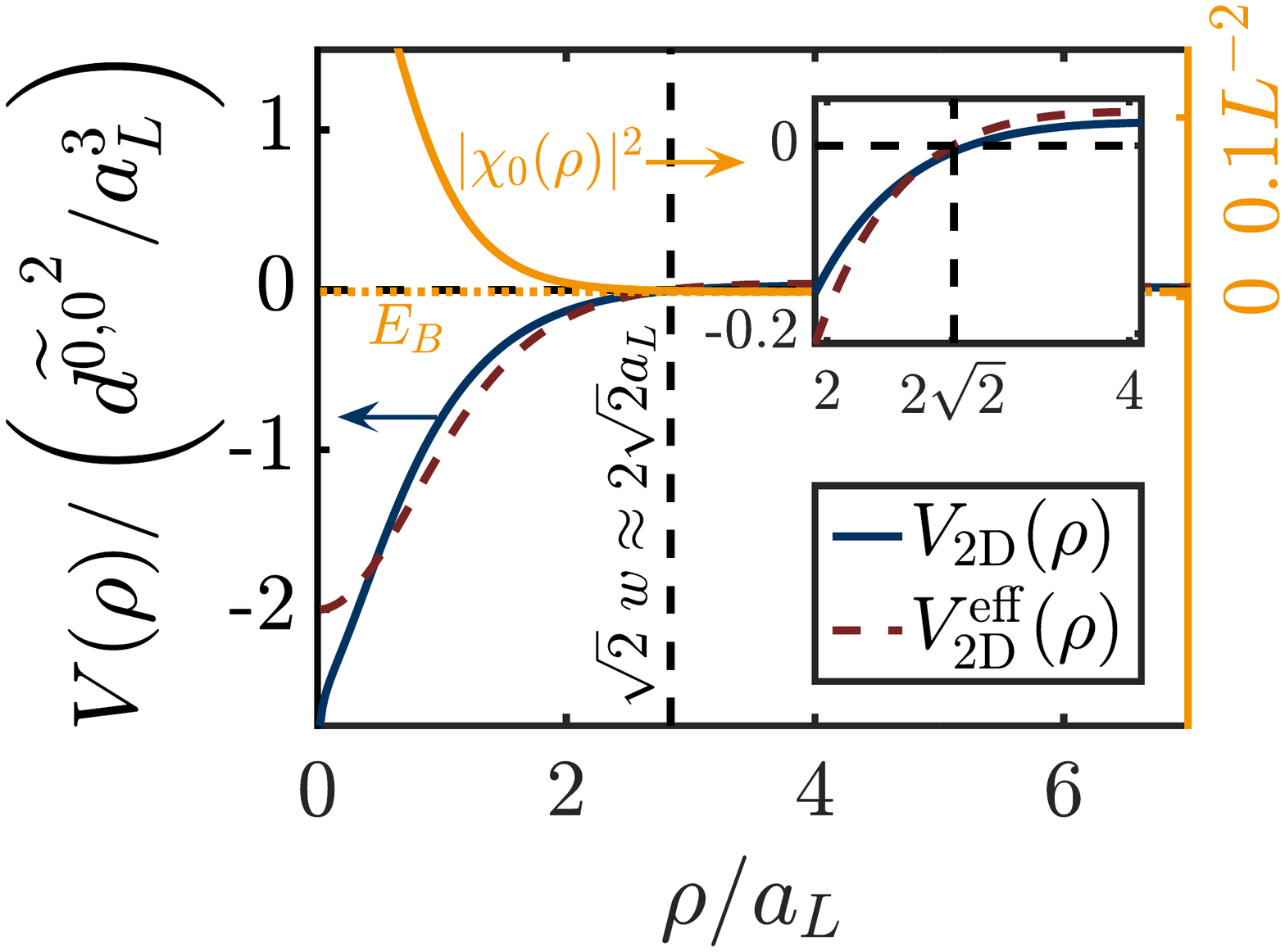}
\llap{\parbox[b]{7.4cm}{(b)\\\rule{0ex}{3.8cm}}}
\vspace{-0.4cm}
\caption{(a) Energy and effective energy of the
  inter-layer bound state as a function of the dipolar length
  $a^{\widetilde 0,0}_d$ for $\epsilon_0 = 3\hbar B/\mathcal{D}$ and
  $l/a_L = 0.1$. The dotted lines indicate that the results for
  $\lvert E_B \rvert > \hbar\omega_L$ should be interpreted as
  order-of-magnitude estimates. For these values of $E_B$, excited
  single-particle levels, which we neglect in our analysis, begin to contribute
  to the bound state. (b) Comparision of the interaction potentials
  $V_\mathrm{2D}(\rho)$ and $V^\mathrm{eff}_\mathrm{2D}(\rho)$. As shown in the
  inset, both potentials change sign at
  $\sqrt{2}~w \approx \sqrt{2}~ 2a_L$. The orange solid line is the
  radial probability density for the bound state $\chi_0(\rho)$, and the dashed
  baseline indicates the corresponding binding energy. Parameters are the same
  as in (a) at $a_d^{\widetilde 0,0}/{a_L = 6.9}$. $L$ denotes the numerical box
  size.}
  
\label{fig:intlambda}
\end{figure}

A solution to Eq.~\eqref{eq:SEEB} with $E_B < 0$ corresponds to a bound
state. In Fig.~\ref{fig:intlambda}(a) and (b), respectively, we show the bound
state energy $E_B$ and the corresponding wave function for different values of
the dipolar length
$a^{\widetilde 0,0}_d = m\,(\mathrm{d}_z^{\widetilde 0,0}/\hbar)^2$.
We  note that the dipolar length scales as
$a_d^{\widetilde 0,0}/a_L\propto \mathcal{D}^2m$
with a prefactor which depends on $l/a_L$ and $\epsilon_0$. The dependence on
$\epsilon_0$ gives an experimental handle to tune the value of the dipolar
length.

\subsection{ Experimental parameters}
\label{sec:LambdaParameters}

\begin{table}[t]
  \centering
  \begin{tabular}{l | c | c | c | c | c | c ||c}
    {} & $\frac{m}{\mathrm{u}}$ & $\frac{\mathcal{D}}{\mathrm{Debye}}$ &
                                                                         $\frac{\hbar B}{h\,\mathrm{GHz}}$\
    & $\frac{\epsilon_0}{ \mathrm{kV}/\mathrm{cm}}$ & $\frac{a^{\widetilde
                                                      0,0}_d}{a_L}$ &
                                                                      $\frac{\lvert E_B \rvert}{h\,\mathrm{kHz}}$&
                                                                      $\frac{\lvert E^{\lambda/2}_B \rvert}{h\,\mathrm{kHz}}$\\ 
    \hline
    KRb\textsuperscript{\cite{de2019degenerate}}   & $127$ & $0.57$ & $ 1.11$ & $ 11.6 $ & $6.9$ & $ 17.6$ & $<0.1$ \\ 
    Toy  & $100$ & $1.00$ & $1.00$ &  6.0  & $16.8$ & $78^*$ & $0.2$ \\ %78
    NaK\textsuperscript{\cite{Park2015, Voges2019}}& $63$ & $2.72$ & $2.83$ & $6.2$ &  $78$ & $> 10^{3^*}$ & $19$\\ 
    LiRb & $91$ & $3.99$ & $7.61$ & $11.4$  & $242$ & $> 10^{3^*}$ & $62$\\ 
    LiCs  & $139$ & $5.39$ & $6.54$ & $7.23$ &  $676$ & $>10^{4^*}$ & $139^*$ 
\end{tabular}
\caption{Parameters for a variety of experimentally relevant fermionic polar
  molecules~\cite{Zuchowski2013, Aymar2005}. For all  molecular species, the
  offset field is chosen as $\epsilon_0 =3.0 \,\hbar B/\mathcal{D}$, which implies
  $\mathrm{d}_z^{\widetilde{0},0} = 0.58\mathcal{D}$, $a_L = 30\,\mathrm{nm}$ and $l/a_L
  = 0.1$. The last column presents the binding energy $E_B^{\lambda/2}$ achievable for a conventional bilayer-stem with inter 
  layer separation $w = \lambda/2 = 250\,\mathrm{nm}$ and an electric offset field of $\epsilon_0 =3.0 \,\hbar B/\mathcal{D}$.
  The asterisk ${}^*$ marks binding energies which exceed the
  single-particle level spacing in the potential well,
  $\lvert E_B \rvert>\hbar\omega_L$, and thus go beyond the range of validity of our analysis which assumes that the molecules populate only the
  two lowest-energy single-particle states. Therefore, these values of $E_B$ should be
  considered as order-of-magnitude estimates.}
  \label{tab:parametersLambda}
\end{table}

In this section we present experimental parameters for the creation of a bilayer system separated on the nanoscale.
As a generic example, we consider
KRb~\cite{de2019degenerate} with mass $m = 127\, \mathrm{u}$ and permanent
molecule-frame dipole moment $\mathcal{D} = 0.57 \, \mathrm{Debye}$. We choose the offset field $\epsilon_0= 3 \, \mathrm{\hbar B/\mathcal{D}}$, which implies
$\mathrm{d}_z^{\widetilde{0},{0}} = 0.58\,\mathcal{D} = 0.33\, \mathrm{Debye}$ and
$a_d^{\widetilde{0},0} = 207\,\mathrm{nm}$.  For $l/a_L = 0.1$, $z_0/a_L = 0.3$
and an harmonic oscillator length of $a_L = 30\,\mathrm{nm}$ the binding energy
is $\lvert E_B \rvert = 2\pi\times18\,\mathrm{kHz} \, \hbar$, well beyond
typical temperatures of
$50\, \mathrm{nK} \, k_{\mathrm{B}} = 2\pi \times 1\,
\mathrm{kHz}\,\hbar$~\cite{de2019degenerate}. In Table~\ref{tab:parametersLambda}
binding energies for different polar molecules are summarized.

From a single-particle point of view, we have to fulfill the hierarchy of scales
$1\gg l/a_L\gg\sqrt{\omega_L/\Omega_c}$ from Eq.~\eqref{eq:inequality}. The
first inequality ensures that the nonadiabatic barrier splits a single potential
well into two sides; The second inequality guarantees stability against nonadiabatic
channel couplings, i.e., $\Gamma_\Lambda/\omega_L\ll1$, see
Eq.~\eqref{eq:Lambda-decay-rate}.  To fulfill the hierarchy of scales we choose
$\Omega_p = 100\, \omega_L$ which leads to
$\Gamma_{\Lambda}/\omega_L < 4 \times 10^{-3}$.  For the sake of
self-consistency we require
$\Gamma_\Lambda < \lvert E_B \rvert = 0.20\,\omega_L$. In absolute numbers, the
Rabi frequencies are given by $\Omega_p = 2\pi\times 8.8\,\mathrm{MHz}$ and
$\Omega_c = 2\pi\times 330\,\mathrm{MHz}$, where we used $\lambda_c = 660\,\mathrm{nm}$~\cite{Wang2010}.
To ensure single-level addressability of rigid rotor eigenlevels it is necessary
to have $\Omega_c < B =2\pi\times 1.1\,\mathrm{GHz}$ for KRb.

We note that with the chosen opposite dipole moments ($\mathrm{d}_z^{\widetilde{0},0}\approx -\mathrm{d}_z^{\widetilde{1},0}/4 $) for the dark state constitutes, the interparticle interaction effectively increases the height of the barrier because of the head-to-head or tail-to-tail orientation of the dipole moments when one of the molecule is under the barrier and the other not. The resulting repulsive dipole-dipole interaction $V$ adds to the barrier potential and, hence, suppresses the wave function amplitude under the barrier and, therefore, the tunneling $J_{L,R}$ and $\Gamma_\Lambda$. The interaction $V$ can be estimated as $V\approx2\mathrm{d}_{z}^{\widetilde{0},{0}}\mathrm{d}_{z}^{\widetilde{1},{0}}/a_{L}^{3}\approx2\pi\times300\,\mathrm{kHz}\,\hbar$ for the case of KRb, which is $\approx0.1\,V_{\mathrm{na}}(0)$. This reduces approximately $J_{L,R}$ by $10\%$  and $\Gamma_\Lambda$ by $5\%$.

We finally compare achievable binding energies for bilayer systems
separated by nanoscale potential barriers to those that can be realized
with standard optical lattices. To do so, we define an effective layer separation as
follows: The interaction potential for dipolar particles which are held fixed at
a relative separation $w$ along the $z$ axis is given
by~\cite{baranov2011bilayer}
\begin{equation}
  V_\mathrm{2D}^\mathrm{eff}(\rho) =
  {\mathrm{d}_z^{\widetilde{0},0}}^2\frac{\rho^2-2w^2}{(\rho^2 +
    w^2)^{5/2}},
\label{eq:EffInt}
\end{equation}
and it changes sign at $\rho = \sqrt{2} w$. As illustrated in
Fig.~\ref{fig:intlambda}(b), also the effective dipolar interaction
$V_\mathrm{2D}(\rho)$ in Eq.~\eqref{eq:V-2D} exhibits a sign change at
$\rho \approx \sqrt{2} 2 a_L$, and we are thus led to identify the effective
layer separation as $w = 2 a_L$. Deviations of
$V_\mathrm{2D}(\rho)$ from the form given in Eq.~\eqref{eq:EffInt} are most
prominent at small separations $\rho$, where $V_\mathrm{2D}(\rho)$ exhibits a
logarithmic divergence due to the nonvanishing overlap of $\psi_L(z)$ and
$\psi_R(z)$. Nevertheless, the direct comparison in Fig.~\ref{fig:intlambda}(a)
shows good quantitative agreement of the binding energies $E_B$ and
$E_B^\mathrm{eff}$ which we obtain for $V_\mathrm{2D}(\rho)$ and
$V_\mathrm{2D}^\mathrm{eff}(\rho)$, respectively.

In conventional bilayer systems which are generated with optical lattices, the
layers are separated by $\lambda/2$, where $\lambda$ is the optical
wavelength~\cite{baranov2011bilayer}.  The inter-layer interaction is then
described by the effective potential $V_\mathrm{2D}^\mathrm{eff}(\rho)$ in
Eq.~\eqref{eq:EffInt}, where $w$ is given by $\lambda/2$. This should be
compared to the value $w = 2 a_L$ which we obtained above for the nanoscale
potential barrier. For $a_d^{\widetilde{0},0}/w\gg 1$, the corresponding binding
energies can be estimated as
$E_B^w \approx 2a^{\widetilde{0},0}_d\hbar^2/(mw^3)$~\cite{baranov2011bilayer}.
Therefore, reducing the effective layer spacing $w$ from
$w = \lambda/2 \gtrsim 250\ \mathrm{nm}$ to achievable harmonic oscillator
lengths of $w = 2 a_L = 60 - 120\,\mathrm{nm}$ increases the binding energy by
1-2 orders of magnitude. Results for a variety of experimentally relevant
molecular species are summarized in Table~\ref{tab:parametersLambda}.

\section{Interface bound state of molecules in electric gradient fields}
\label{sec:2LS}

We now consider the setup illustrated in Fig.~\ref{fig:setup}, in which the
electric field gradient created by a standing optical wave is replaced by a
static electric gradient field. Further, in contrast to the $\Lambda$-systems
with optical transitions which we considered in the previous section, we focus
now on effective two-level systems, which are formed by two rotational levels
that are coupled by MW fields. As we show, in the adiabatic limit, when the
single-particle dynamics occurs in a single BO channel, bound states of two
molecules can form at the interface at which the induced dipole moment changes
sign.

\subsection{Single-particle physics}
\label{sec:ssp2LS}

\begin{figure}[t]
\centering
\vspace{-0.8cm}
\hspace{0.3cm}
\includegraphics[clip,trim=6.5cm 13.5cm 10.cm 7.3cm,height=.5\linewidth]{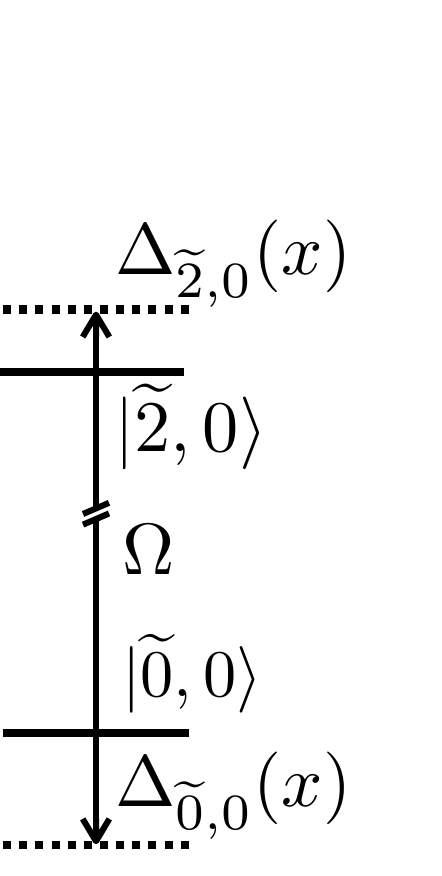}
\llap{\parbox[b]{5.2cm}{(a)\\\rule{0ex}{2.9cm}}}
\includegraphics[clip,trim=6cm 10.5cm 5cm 10cm,height=.45\linewidth]{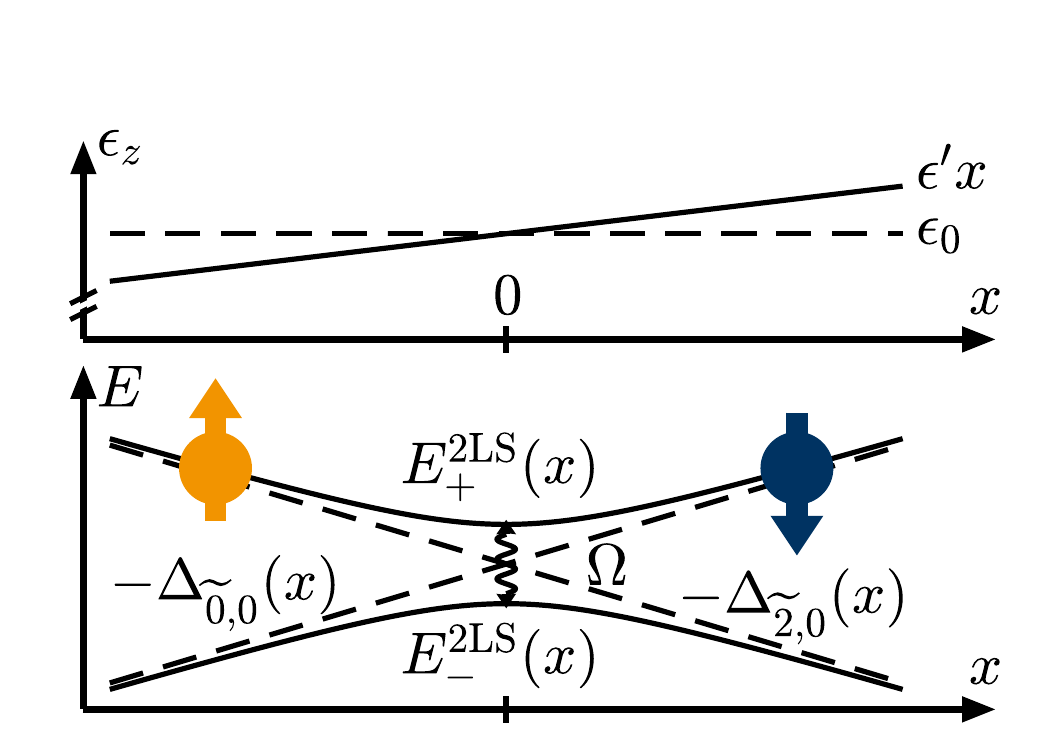}
\llap{\parbox[b]{11.2cm}{(b)\\\rule{0ex}{2.9cm}}}
\llap{\parbox[b]{11.4cm}{(c)\\\rule{0ex}{1.6cm}}}
\caption{(a) The two rotational states
  $\ket{\widetilde{0},0}$ and $\ket{\widetilde{2},0}$ are coupled by a microwave with Rabi frequency $\Omega$. (b) The detunings $\Delta_{\widetilde N,0}(x)$, where $N = 0,2$ are position dependent due to a spatially non-uniform static electric field $\epsilon(x) = \epsilon_0+\epsilon'x$. (c) Therefore, the micro wave coupling is resonant with the position-dependent Stark shifted energies  (dashed lines) of two rotational states with opposite induced dipole moments (blue and orange arrows) at a particular value of the x coordinate which we choose as the origin of the coordinate system. This gives rise to the two dressed channels with energies $E_\pm^\mathrm{2LS}(x)$.}
\label{fig:2LS-configuration}
\end{figure}

The key elements to engineer interface bound states are a position-dependent
electric field, $\be(x) = \be_0 + \be'x$, which comprises both a homogeneous
offset field $\be_0$ as discussed at the beginning of
Sec.~\ref{sec:LambdaSystem} and a gradient field
$\be' x = \epsilon' \mathbf{e}_z x$, and a MW-induced Rabi coupling $\Omega$ of
the molecular levels $\ket{\widetilde{0}, 0}$ and $\ket{\widetilde{2}, 0}$, see Fig.~\ref{fig:2LS-configuration}(a) and (b). 
Note, that these states can be coupled by MW field because of the admixture of the state $\ket{1,0}$ in both of them for $\epsilon_0\neq 0$. 
The reason for the choice of these internal states will be clarified below. As illustrated in
Fig.~\ref{fig:stark}(b), the strong offset field $\be_0$ induces state-dependent
dipole moments $\bd^{\widetilde N, m}$, which couple to the gradient field
$\be' x$ and thus give rise to position-dependent Stark shifts. To calculate
these Stark shifts, we note that in the present setup the motion of the
molecules along the $x$-direction is restricted to a region of spatial extent
$l_0$ which we determine below to be on the order of tens of nanometers, and we
assume $\epsilon_0\gg\epsilon' l_0$, so that the gradient field can be taken
into account perturbatively. Then, to first order in the gradient field, the
Stark shift of the rotational energy levels is
$E_{\widetilde{N},m}[\be_0 + \be' x] \approx E_{\widetilde{N},m}(\be_0) -
[\mathbf{d}^{\widetilde{N},m}(\be_0) \cdot \be']\, x$.
To simplify the notation, we omit the dependency on $\be_0$ in the following.
Higher orders in the expansion in $\be' x$ are negligible for the large electric
offset fields considered in the following sections. As a result, the molecules
move in a state-dependent linear potential, and the Rabi coupling between the
internal states is resonant only at a particular point, which we chose as the
coordinate origin. The Hamiltonian which describes the motion of a single
molecule in this configuration of electric and MW fields is given by
\begin{equation}
  \label{eq:H-2LS}
  H_{\mathrm{2LS}} = \frac{p_x^2}{2m} + H_{\mathrm{2LS}}^0(x),
\end{equation}
where $p_x = -i\hbar\partial_x$ is the $x$-component of the momentum
operator. In a proper rotating frame, the MW coupling Hamiltonian reads
\begin{equation}
  H_{\mathrm{2LS}}^0(x) = \hbar\begin{pmatrix} -\Delta_{\widetilde 2, 0}(x) & \Omega/2 \\ \Omega/2 &
    -\Delta_{\widetilde 0, 0}(x) \end{pmatrix}, 
      \label{eq:2LSHam}
\end{equation}
with linearly position-dependent detunings,
$\Delta_{\widetilde N, 0}(x) = \Delta_{\widetilde N, 0}' x$, where
$\Delta_{\widetilde N, 0}' = -\bd^{\widetilde N, 0}\cdot\be'/\hbar$. Due to its
dependence on the position $x$, the Hamiltonian $H_{\mathrm{2LS}}^0(x)$ couples
internal and external degrees of freedom. Typically, the energy scales
associated with the internal degrees of freedom are much larger than those which
characterize the motion of molecules. This allows us to treat the internal
dynamics in a BO approximation. To wit, we first diagonalize the 2LS
Hamiltonian~\eqref{eq:2LSHam}; Thereby, we treat the position $x$ as a
parameter. The eigenvalues $E^\mathrm{2LS}_{\pm}(x)$ of $H^0_{\mathrm{2LS}}(x)$ are given by
\begin{equation}
  E^\mathrm{2LS}_\pm(x) = \frac{\hbar\Omega}{2}\left[\frac{1-\delta}{1+\delta} \frac{x}{s} \pm\sqrt{1+\left(x/s \right) ^2}\right],
  \label{eq:Epm}
\end{equation}
as illustrated in  Fig.~\ref{fig:2LS-configuration}(c), where $-\delta = \Delta'_{\widetilde 0,0}/\Delta'_{\widetilde 2,0} =
\mathrm{d}^{\widetilde 0,0}_z/ \mathrm{d}^{\widetilde 2,0}_z$
is the ratio of induced dipole moments in the states $\ket{\widetilde{0}, 0}$
and $\ket{\widetilde{2}, 0}$, and
\begin{equation}
\label{eq:resarea}
s = \frac{\Omega}{\Delta'_{\widetilde 2,0}(1+\delta)}
\end{equation}
corresponds to the length scale which delimits the spatial region in which the
Rabi coupling is resonant. 
The corresponding eigenvectors read
\begin{equation}
  \label{eq:plus-minus-channel-states}
  \begin{split}
    \ket{+}_x & = \sin(\theta_x/2)\ket{\widetilde 2, 0} +
    \cos(\theta_x/2)\ket{\widetilde 0, 0}, \\ \ket{-}_x & =
    \cos(\theta_x/2)\ket{\widetilde 2, 0} - \sin(\theta_x/2)\ket{\widetilde 0,
      0},
  \end{split}
\end{equation}
where $\theta_x = \mathrm{atan}(s/x)$ and $0 \leq \theta_x \leq \pi$. The states
$\ket{\pm}_x$ are superpositions of the rotational states
$\ket{\widetilde{0}, 0}$ and $\ket{\widetilde{2}, 0}$ with spatially varying
amplitudes. In particular, $\ket{+}_x \to \ket{\widetilde{0}, 0}$ and
$\ket{+}_x \to \ket{\widetilde{2}, 0}$ for $x \to + \infty$ and
$x \to - \infty$, respectively, with the transition occurring in a spatial
region of extent $s$. Correspondingly, the limiting behavior of $\ket{-}_x$ is
given by $\ket{-}_x \to \ket{\widetilde{2}, 0}$ and
$\ket{-}_x \to \ket{\widetilde{0}, 0}$ for $x \to + \infty$ and
$x \to - \infty$, respectively.

As already discussed in Sec.~\ref{sec:Lambda} the BO approximation
\cite{burrows2017nonadiabatic, kazantsev1990mechanical, dum1996gauge,
  dutta1999tunneling, ruseckas2005non} assumes that the internal state of the
molecules adiabatically follows the external motion, where the parametric
dependence of the internal state on the position is given by
Eq.~\eqref{eq:plus-minus-channel-states}. In particular, the Hamiltonian for the
motion of a molecule prepared in the $+$ channel is given by
\begin{equation}
  \label{eq:+-BO-channel}
  H_{\mathrm{2LS}, ++}^{\mathrm{ad}} = \frac{p_x^2}{2 m} + E^\mathrm{2LS}_+(x) +
  V_{\mathrm{na}}(x),
\end{equation}
where the nonadiabatic potential barrier is given by 
\begin{equation}
  \label{eq:2LS-na-potential-barrier}
  V_{\mathrm{na}}(x) = 
  \frac{\hbar^2}{2m s^2}\frac{1}{4} \frac{1}{[1+(x/s)^2]^2}.
\end{equation}
A detailed derivation of the adiabatic Hamiltonian from Eq.~\eqref{eq:+-BO-channel} is provided in Appendix \ref{sec:basistrans}.
We focus on a parameter regime that is specified below, in which both the
channel coupling and the nonadiabatic contribution to the effective BO potential
from $A^2(x)$ are negligible. The validity of this approximation is discussed in
detail in Appendix~\ref{sec:decayBO}.

The effective BO potential corresponding to the eigenvalue $E_+(x)$ forms a
trap. This is because the induced dipole moments $\mathrm{d}^{\widetilde 2,0}_z$
and $\mathrm{d}^{\widetilde 0,0}_z$ have opposite sign as can be seen in
Fig.~\ref{fig:stark}(b), and thus the position-dependent Stark shifts
$\Delta_{\widetilde{0}, 0}(x)$ for $x \gtrsim s$ and
$\Delta_{\widetilde{2}, 0}(x)$ for $x \lesssim -s$ have opposite signs. For
$\lvert x \rvert \ll s$ we approximate the BO trapping potential $E^\mathrm{2LS}_+(x)$ by a
harmonic potential,
\begin{equation}
  \label{eq:harmonic-approx-+-channel}
  \begin{split}
    E^\mathrm{2LS}_+(x) & \approx\hbar\Omega/2\left[1-x_0^2/s^2+(x-x_0)^2/s^2\right] \\ & =
    \Delta E + m \omega_0^2(x-x_0)^2/2.
  \end{split}
\end{equation}
The harmonic approximation is characterized by the effective trap frequency
$\hbar\omega_0 = \sqrt{\hbar\Omega~\hbar^2/(2ms^2)}$, the position of the
minimum $x_0/s = (1-\delta)/2\sqrt \delta$, and the energy shift
$\Delta E = \hbar \Omega/2[1-(x_0/s)^2]$. In this harmonic potential, the size
of the ground state is given by $l_0 = \sqrt{\hbar/(m\omega_0)}$, and for
self-consistency we require $l_0/s \ll 1$. The effective potential $E_-(x)$ in
the BO channel corresponding to the eigenstate $\ket{-}_x$ forms an inverted
trap, and consequently this channel hosts a continuum of scattering states. The
coupling between the BO channels leads to decay of the bound states in the
$+$ channel to the continuum in the $-$ channel. The decay rate can be estimated
by Fermi's golden rule as
\begin{equation}
  \frac{\Gamma_\mathrm{2LS}}{\omega_0} \approx 2\sqrt{\pi}\frac{l_0}{s}\exp\left[-8\left(\frac{s}{l_0} \right) ^2 \right].
\label{eq:2LSlimit}
\end{equation}
A detailed derivation is provided in Appendix~\ref{sec:decayBO}. The adiabatic
limit requires $\Gamma_\mathrm{2LS}/\omega_0 \ll 1$ which is achieved for
$l_0/s \ll 1$, in agreement with the harmonic approximation of $E_+(x)$. Because
of the exponential factor in the rate Eq.~\eqref{eq:2LSlimit} a moderately small
ratio of $l_0/s =1/\sqrt{2}$ is sufficient to obtain a strongly suppressed decay
rate of $\Gamma_{\mathrm{2LS}}/\omega_0 < 10^{-6}$. In the next section, we
focus on the $+$ channel and discuss bound states of pairs of molecules due to
spatially inhomogeneous dipole moments.

\subsection{Dipolar interaction and interface bound state}
\label{sec:2LS-bound-state}

Since the state $\ket{+}_x$ defined in Eq.~\eqref{eq:plus-minus-channel-states}
changes from $\ket{+}_x \to \ket{\widetilde{0}, 0}$ for $x \to + \infty$ to
$\ket{+}_x \to \ket{\widetilde{2}, 0}$ for $x \to - \infty$, also the induced
dipole moment $\mathrm{d}^{\mathrm{D}}_z(x)$ defined in
Eq.~\eqref{eq:induced-dipole-moment-2LS} below varies in space and takes
limiting values with opposite signs given by
$\mathrm{d}^{\mathrm{D}}_z(x) \to \mathrm{d}_z^{\widetilde{0}, 0}$ and
$\mathrm{d}^{\mathrm{D}}_z(x) \to \mathrm{d}_z^{\widetilde{2}, 0}$ for
$x \to + \infty$ and $x \to - \infty$, respectively. Therefore, two molecules in
the $+$ channel experience attractive induced dipolar interactions if they
are located on opposite sides of the point $x_0$ where the dipole moment
$\mathrm{d}_z(x)$ changes sign. If they are on the same side, they repel each other~\cite{note:hyperfine}.
As we show in the following, this gives rise to
the formation of bound states of two molecules close to $x_0$.

We consider now the full 3D geometry with a harmonic confinement
$V_{\perp}(z) = m \omega_{\perp}^2 z^2/2$ in the $z$ direction, whereas motion
along the $y$ axis is unrestricted. The Hamiltonian for the motion of two
molecules $j = 1,2$ with position coordinates $\br_j = (x_j, y_j, z_j)$ and
momenta $\mathbf{p}_j$ is then given by
\begin{align}
  \label{eq:H-two-body-2LS}
  &H_{2, \mathrm{2LS}} = \sum_{j=1,2} \left[ \frac{\mathbf{p}_j^2}{2m} +
    H^0_{\mathrm{2LS}}(x_j) + \frac{1}{2} m\omega_{\perp}^2 z_j^2 \right] \\ 
    &+U^{(1)}_{\mathrm{2LS}}(t)\otimes U^{(2)}_{\mathrm{2LS}}(t) ~
  V_{\mathrm{dd}}(\br_1 - \br_2) ~ U_{\mathrm{2LS}}^{(2),\dagger}(t)\otimes
  {U^{(1),\dagger}_{\mathrm{2LS}}}(t),\notag
\end{align}
where the dipolar interaction is given in Eq.~\eqref{eq:V-dd} and the
rotating-frame transformation
$U^{(j)}_{\mathrm{2LS}}(t) = \exp\left[ -i(E_{\widetilde 2, 0} - E_{\widetilde
    0,0}) t/\hbar\,\ket{\widetilde 0,0}\bra{\widetilde 0,0} \right]$
for the $j$\textsuperscript{th} particle.  The wave function for two molecules
in the $+$ BO channel can be written as
\begin{equation}
  \label{eq:two-body-wave-function-2LS}
 \ket{\Psi} = \int \mathrm{d}\br_1\mathrm{d}\br_2 \,\Psi(\br_1,\br_2)
 \ket{\br_1,\br_2}\otimes\ket{+}_{x_1}\otimes\ket{+}_{x_2}.
\end{equation}
In the following, we neglect nonadiabatic corrections to the BO
approximation. Upon projecting the Hamiltonian in Eq.~\eqref{eq:H-two-body-2LS}
to the $+$ channel, the contribution $H_{\mathrm{2LS}}^0(x_j)$ is replaced by
the effective potential $E_+(x_j)$. Further, the dipolar interaction for
two molecules in the $+$ channel is given by
\begin{multline}
  V_{\mathrm{dd}}^+(\mathbf{r}_1, \mathbf{r}_2) = {}_{x_1}\bra{+}\otimes{}_{x_2}\bra{+}
   \left[ U^{(1)}_{\mathrm{2LS}}(t)\otimes U^{(2)}_{\mathrm{2LS}}(t) \right. \\
   \left. \times V_{\mathrm{dd}}(\br_1 - \br_2) \,
  U_{\mathrm{2LS}}^{{(2)},\dagger}(t) \otimes
  U_{\mathrm{2LS}}^{{(1)},\dagger}(t) \right] \ket{+}_{x_1}\otimes\ket{+}_{x_2}.
\end{multline}
Due to the rotating-frame transformation $U_{\mathrm{2LS}}^{(j)}(t)$, certain
contributions to the dipolar interaction acquire rapidly oscillating
phase factors. These contributions average to zero, and we only keep the
time-independent components. Then, the dipolar interaction can be written
as
% \begin{equation}
%   \label{eq:V-dd-+-channel}
%   \begin{split}
%     V_{\mathrm{dd}}^+(\mathbf{r}_1, \mathbf{r}_2) &=
%     \frac{\mathrm{d}^{\mathrm{D}}_z(x_1)\mathrm{d}^{\mathrm{D}}_z(x_2)+2\mathrm{d}^{\mathrm{E}}_z(x_1)\mathrm{d}^{\mathrm{E}}_z(x_2)}{|\br_1-\br_2|^3}
%     \\ & \mathrel{\hphantom{=}} \times \left[1-3\frac{(z_1-z_2)^2}{|\br_1-\br_2|^2} \right]\\
%     &\approx
%     \frac{\mathrm{d}^{\mathrm{D}}_z(x_1)\mathrm{d}^{\mathrm{D}}_z(x_2)}{|\br_1-\br_2|^3}
%     \left[1-3\frac{(z_1-z_2)^2}{|\br_1-\br_2|^2} \right],
%   \end{split}
% \end{equation}
\begin{multline}
  \label{eq:V-dd-+-channel}  
  V_{\mathrm{dd}}^+(\mathbf{r}_1, \mathbf{r}_2) =
  \frac{\mathrm{d}^{\mathrm{D}}_z(x_1)\mathrm{d}^{\mathrm{D}}_z(x_2)+2\mathrm{d}^{\mathrm{E}}_z(x_1)\mathrm{d}^{\mathrm{E}}_z(x_2)}{|\br_1-\br_2|^3}
  \\ \times \left[1-3\frac{(z_1-z_2)^2}{|\br_1-\br_2|^2} \right],  
\end{multline}
where the ``direct'' dipole moments are
\begin{equation}
  \label{eq:induced-dipole-moment-2LS}
  \begin{split}
    \mathrm{d}^{\mathrm{D}}_z(x) & = {}_x\bra{+} \left(\ket{\widetilde
        2,0}\bra{\widetilde 2,0} \hat{\mathrm{d}}_z \ket{\widetilde
        2,0}\bra{\widetilde 2,0} \right. \\ & \left. \mathrel{\hphantom{=}} +
      \ket{\widetilde 0,0}\bra{\widetilde 0,0}\hat{\mathrm{d}}_z \ket{\widetilde
        0,0}\bra{\widetilde 0,0}\right)\ket{+}_x \\ & = \mathrm{d}_z^{\widetilde
      2, 0} \left(\frac{1-\delta}{2}-\frac{1+\delta}{2}\frac{x}{\sqrt{s^2+x^2}}
    \right),
  \end{split}
\end{equation}
and the ``exchange'' moments, which correspond to interaction processes that
exchange the internal states of the molecules, read
\begin{equation}
  \begin{split}
    \mathrm{d}^{\mathrm{E}}_z(x) & = {}_x\bra{+}{\widetilde
      2,0}\rangle\bra{\widetilde 2,0} \hat{\mathrm{d}}_z \ket{\widetilde
      0,0}\langle{\widetilde 0,0}\ket{+}_x \\ & =\bra{\widetilde 2,0}
    \hat{\mathrm{d}}_z \ket{\widetilde 0,0}\frac{1}{2} \frac{s}{\sqrt{s^2+x^2}}.
  \end{split}
\end{equation}
In the following, we neglect contributions to $V_\mathrm{dd}^+(\br_1,\br_2)$
which involve $\mathrm{d}^{\mathrm{E}}_z(x)$. This is justified since
$|\bra{\widetilde 2,0} \hat{\mathrm{d}}_z \ket{\widetilde
  0,0}|\ll|\mathrm{d}_z^{\widetilde 2,0}|,~|\mathrm{d}_z^{\widetilde 0,0}|$.
We note that a corresponding relation does not apply, for example, for the pair
of states $\ket{\widetilde 0, 0}$ and $\ket{\widetilde 1, 0}$. We further remark
that the diagonal matrix elements of the dipole moment operator which enter
Eq.~\eqref{eq:induced-dipole-moment-2LS} are not affected by the rotating-frame
transformation which lead to the 2LS Hamiltonian~\eqref{eq:2LSHam}. The position
$x_0$ of the zero-crossing of $\mathrm{d}_z^{\mathrm{D}}(x)$ coincides with the
minimum of $E_+(x)$ from Eq.~\eqref{eq:Epm} given above. For $l_0 \ll s$ we expand
\begin{equation}
  \mathrm{d}_z^{\mathrm{D}}(x_0 + x) \approx - \mathrm{d}_z^{\widetilde{2}, 0}
  \frac{4\delta^{3/2}}{(1+\delta)^2}\frac{x}{s}
  \label{eq:redDip}
\end{equation}
to linear order in $x/s$ around $x_0$. Here and in the following, the
$x$-coordinates $x_{1,2}$ of the molecules are measured from the zero-crossing
$x_0$ of the induced dipole moment~\eqref{eq:induced-dipole-moment-2LS}. Then,
the projection of the two-molecule Hamiltonian~\eqref{eq:H-two-body-2LS} to the
$+$ channel reads 
\begin{equation}
  \label{eq:H-two-body-2LS-+-channel}
  H_{2, \mathrm{2LS}}^+ = \sum_{j = 1,2} \left[ \frac{\mathbf{p}_j^2}{2 m} +
    E^\mathrm{2LS}_+(x_j) +
    \frac{1}{2} m \omega_{\perp}^2 z_j^2
  \right] + V_{\mathrm{dd}}^+(\mathbf{r}_1, \mathbf{r}_2).
\end{equation}
Since the induced dipole moment~\eqref{eq:induced-dipole-moment-2LS} is oriented
along the $z$ axis, inelastic head-to-tail collisions of the two molecules can
be suppressed through tight confinement in this direction. We assume that the
trapping frequency $\omega_{\perp}$ is sufficiently large such that excited
states in the potential $V_{\perp}(z)$ are energetically inaccessible, and the
molecules reside in the ground state. Then, the wave function
$\Psi(\br_1,\br_2)$ in Eq.~\eqref{eq:two-body-wave-function-2LS} factorizes as
$\Psi(\br_1,\br_2) = \Psi_{\parallel}(\boldsymbol{\rho}_1, \boldsymbol{\rho}_2)
\, \phi^0_{\perp}(z_1) \, \phi^0_{\perp}(z_2)$
where $\boldsymbol{\rho}_j = \left( x_j, y_j \right)$. The harmonic oscillator
ground state wave function reads
\begin{equation}
  \phi^0_{\perp}(z) = \frac{1}{\sqrt{l_{\perp}\sqrt{\pi}}} \exp \left(-\frac{z^2}{2l_{\perp}^2} \right) ,
\end{equation}
where $l_{\perp}^2 = \hbar/(m \omega_{\perp})$ is the corresponding oscillator
length. Under these conditions, which also imply $l_{\perp} \ll l_0$, the motion
of the molecules is confined to the $xy$ plane, and the system becomes
effectively 2D. Up to the zero-point energy in the harmonic confinement in the
$z$ direction, the Hamiltonian for the 2D motion of the two molecules is given
by
\begin{equation}
  \label{eq:H-2D-2LS}
  H_{\mathrm{2D}} = \sum_{j = 1,2} \left[ \frac{\mathbf{p}_j^2}{2 m} +
    E^\mathrm{2LS}_+(x_j) \right] + V_{\mathrm{2D}}(\boldsymbol{\rho}_1,
  \boldsymbol{\rho}_2).
\end{equation}
We obtain the effective 2D dipolar interaction by integrating out the
tightly confined $z$ direction, which yields
\begin{equation}
  \label{eq:V2D}
  \begin{split}
    V_{\mathrm{2D}}(\boldsymbol{\rho}_1, \boldsymbol{\rho}_2) &= \int
    \mathrm{d}z_1 \mathrm{d}z_2 \, \lvert \phi^0_{\perp} (z_1) \rvert^2 \,
    \lvert \phi^0_{\perp}
    (z_2) \rvert^2 \, V_{\mathrm{dd}}^+(\mathbf{r}_1, \mathbf{r}_2) \\
    & = \mathrm{d}_z^{\mathrm{D}}(x_1) \mathrm{d}_z^{\mathrm{D}}(x_2)
    v_{\mathrm{2D}}(\rho).
  \end{split}
\end{equation}
That is, the effective 2D dipolar interaction can be written as the
product of position-dependent dipole moments $\mathrm{d}_z^{\mathrm{D}}(x_j)$
and an interaction potential $v_{\mathrm{2D}}(\rho)$ which depends only on the
relative distance in the $xy$ plane, $\rho^2 = (x_1 - x_2)^2 + r_y^2$, where
$r_y = y_1 - y_2$ is the relative coordinate in the $y$ direction. The
interaction potential is given by
\begin{multline}
  v_{\mathrm{2D}}(\rho) = \frac{1}{\sqrt{8\pi}l_{\perp}^3} \exp \!
  \left(\frac{\rho^2}{4l_{\perp}^2} \right) \left[
    \left(2+\frac{\rho^2}{l_{\perp}^2} \right) K_0 \!
    \left(\frac{\rho^2}{4l_{\perp}^2} \right) \right. \\
  \left. -\frac{\rho^2}{l_{\perp}^2} K_1 \!  \left(\frac{\rho^2}{4l_{\perp}^2}
    \right) \right].
\end{multline}
Here, $K_n(z)$ is the modified Bessel function of the second kind. At large
distances $\rho \gg l_{\perp}$, the interaction potential assumes the
characteristic dipolar form $v_{\mathrm{2D}}(\rho) \sim 1/\rho^3$; At short
distances $\rho \ll l_{\perp}$ it diverges logarithmically,
$v_{\mathrm{2D}}(\rho) \sim \sqrt{2/\pi}/l_\perp^3\ln(\rho/l_{\perp})$. 

The setup we consider is translationally invariant along the $y$ axis and,
therefore, the motion of two molecules in this direction factorizes into
center-of-mass (COM) and relative components,
$\Psi_{\parallel}(\boldsymbol{\rho}_1, \boldsymbol{\rho}_2) = \psi(x_1, x_2,
r_y) \, \xi(R_y)$,
where $R_y = (y_1 + y_2)/2$ is the COM coordinate. Possible bound states are
negative-energy solutions of the two-body SE
\begin{multline}
  \label{eq:two-body-SE}
  \left[ \sum_{j=1,2} \left( -\frac{\hbar^2}{2m}\frac{\partial^2}{\partial
        x_j^2}+\frac{1}{2}m\omega_0^2 x_j^2 \right)
    -\frac{\hbar^2}{m}\frac{\partial^2}{\partial r_y^2} \right. \\
  \left. \vphantom{\sum_{j=1,2} \left(
        -\frac{\hbar^2}{2m}\frac{\partial^2}{\partial
          x_j^2}+\frac{1}{2}m\omega_0^2 x_j^2 \right)} +
    V_{\mathrm{2D}}(x_1,x_2, r_y) \right] \psi(x_1,x_2,r_y) \\ = \left(
    E_B+\hbar\omega_0 \right) \psi(x_1,x_2,r_y),
\end{multline}
where we replaced the effective potential $E_+(x)$ in the $+$ BO channel by its
harmonic approximation~\eqref{eq:harmonic-approx-+-channel}, and we omitted the
energy offset $\Delta E$. The binding energy $E_B$ is measured from the ground
state energy of the noninteracting two particle problem, which is
$2\times \hbar\omega_0/2$.

\begin{figure}[t]
\centering
\hspace{-0.3cm}
\includegraphics[clip,trim=2.5cm 6.5cm 5.cm 7.cm,height=.22\textwidth]{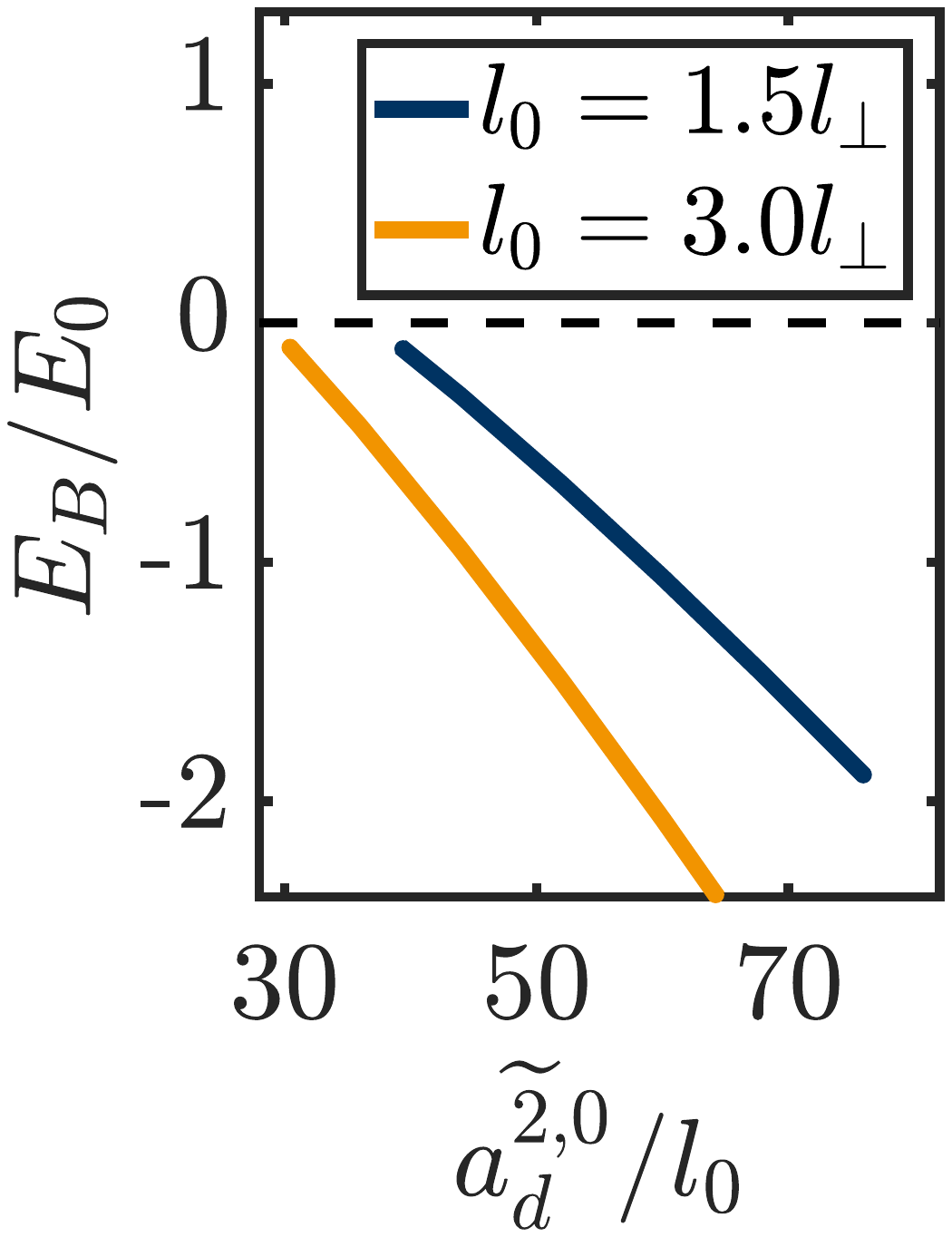}
\llap{\parbox[b]{4.9cm}{(a)\\\rule{0ex}{4cm}}}
\hspace{-0.4cm}
\includegraphics[clip,trim=1.7cm 5.9cm 3cm 7.1cm,height=.22\textwidth]{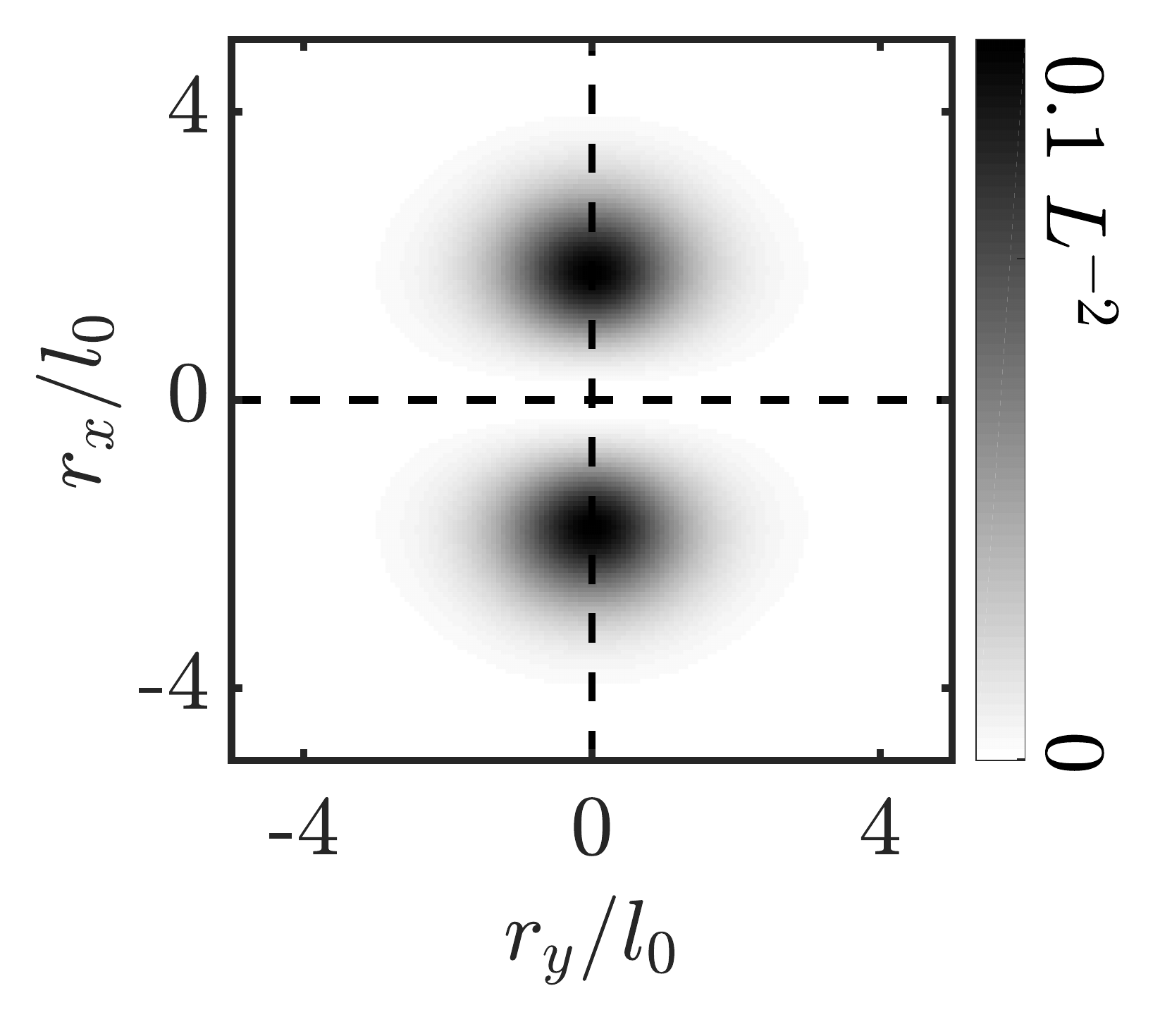}
\llap{\parbox[b]{7.cm}{(b)\\\rule{0ex}{4cm}}}
\includegraphics[clip,trim=1.8cm 7cm 2.5cm 11.05cm,height=.23\textwidth]{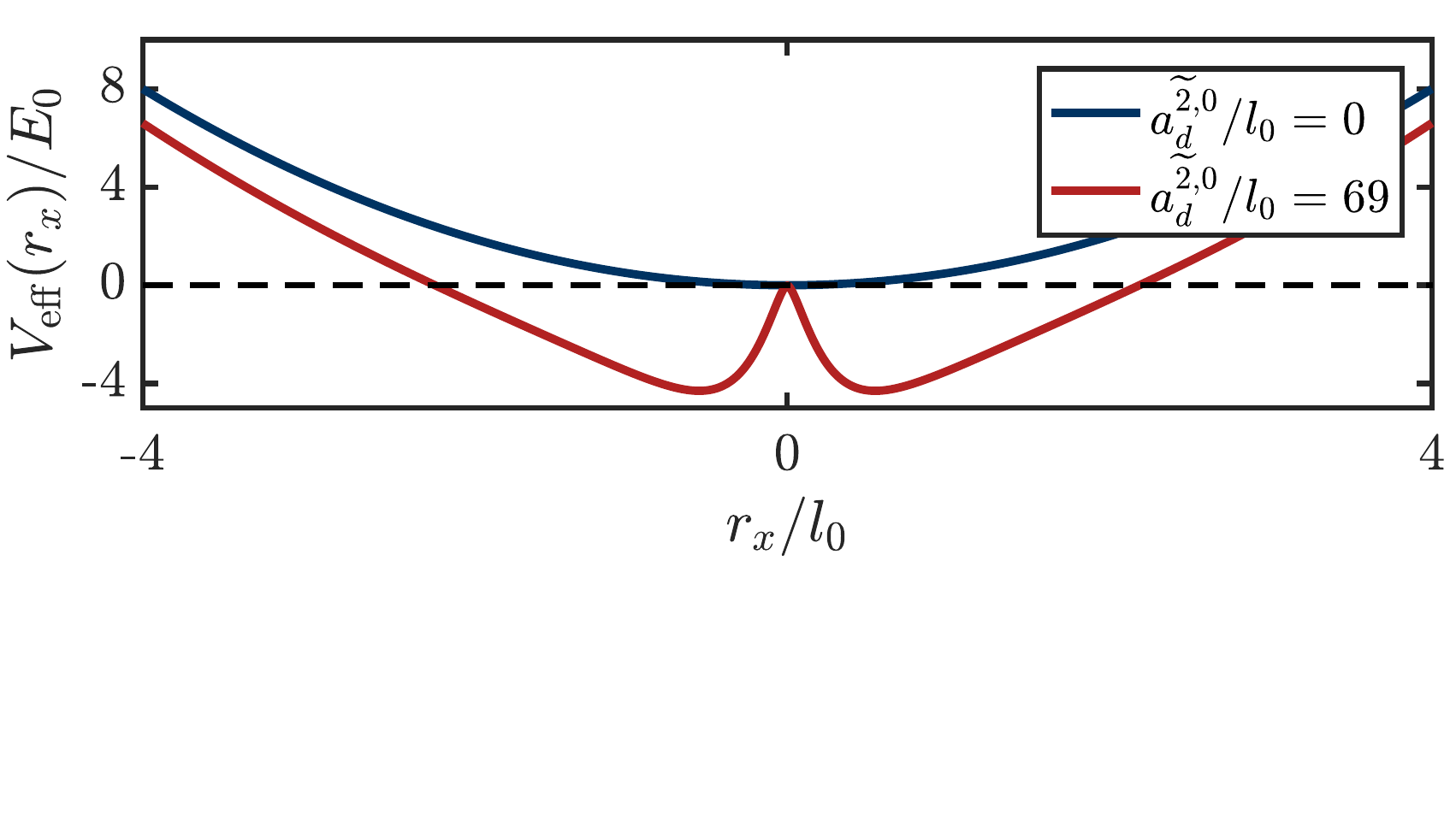}
\llap{\parbox[b]{12.9cm}{(c)\\\rule{0ex}{4cm}}}
\vspace{-1.7cm}
\caption{(a) Energy of the bound state in units of $E_0 = \hbar^2/2ml_0^2$ as a
  function of the dipolar length $a^{\widetilde 2,0}_d$ defined in
  Eq.~\eqref{eq:dipolar-length} for several values of $l_0/l_\perp$ and
  $s = \sqrt{2} l_0$. (b) Probability density
  $\int \mathrm{d}R_x|\psi(r_x,R_x,r_y)|^2$ of the bound state, where
  $r_x = x_1 - x_2$ and $R_x = (x_1 + x_2)/2$ are the relative and COM
  coordinates in the $x$ direction, for $l_0 = 1.5l_\perp$ and $a_d/l_{0} = 69$.
  $L$ denotes the box size used in the numerical diagonalization of
  Eq.~\eqref{eq:two-body-SE}. (c) Effective potential~\eqref{eq:V-eff} for the
  simplified model discussed in the main text.}
\label{fig:tls}
\end{figure}

We solve this Eq.~\eqref{eq:two-body-SE} numerically. Details are discussed in
Appendix~\ref{sec:interface-bound-state-numerics}, and we present our results in
Fig.~\ref{fig:tls}. As shown in Fig.~\ref{fig:tls}(a), a bound state occurs for
sufficiently strong induced dipolar interactions, where the strength of dipolar
interactions is characterized by the ratio $a^{\widetilde 2,0}_d/l_0$ with
\begin{equation}
  \label{eq:dipolar-length}
  a^{\widetilde 2,0}_d = m\left[\frac{\mathrm{d}_z^{\widetilde{2}, 0}}{\hbar}
    \frac{4\delta^{3/2}}{(1+\delta)^2}\right]^2.
\end{equation}  
We note that while achievable binding energies are boosted if the characteristic
length scale $l_0$ of this setup is on the order of tens of nanometers, the
scheme suffers from a reduction of the dipolar moment in the vicinity of the
interface. Figure~\ref{fig:tls}(b) shows the wave function of the bound state
for $l_0 = 1.5l_\perp$ and $a_d^{\widetilde 2,0}=69l_0$.  Numerically, we find
that the threshold value for the formation of a bound state depends only
slightly on $l_\perp$. This is because $V_{\mathrm{2D}}(x_1,x_2, r_y)$ is only
modified in regions where $(x_1-x_2)^2+r_y^2\lesssim l_\perp^2$, and in these
regions, the probability amplitude $|\psi(x_1,x_2,r_y)|^2$ is suppressed. The
existence of a threshold value of the ratio $a^{\widetilde 2,0}_d/l_0$ implies
that strong harmonic confinement suppresses the formation of the bound state.

These results can be understood qualitatively within a simplified model which we
obtain from Eq.~\eqref{eq:two-body-SE} by setting both the COM coordinate in the
$x$ direction, $R_x = (x_1 + x_2)/2$, and the relative coordinate in the $y$
direction to zero, $R_x = r_y = 0$. This yields an effective SE for the
component of the wave function which describes the relative motion of the two
molecules in the $x$ direction. The corresponding effective potential, which
depends only on the relative coordinate $r_x = x_1 - x_2$, contains
contributions from the harmonic confinement and the dipolar interaction and is
given by
\begin{equation}
  \label{eq:V-eff}
  V_{\mathrm{eff}}(r_x) = \frac{1}{4} m \omega_0^2 r_x^2 +
  V_{\mathrm{2D}}(r_x/2, - r_x/2, 0).
\end{equation}
Figure~\ref{fig:tls} shows the effective potential for
$a_d^{\widetilde 2,0}/l_0 = 0$ and $a_d^{\widetilde 2,0}/l_0 = 69$. In the
former case, $V_{\mathrm{eff}}(r_x)$ reduces to a harmonic potential. Then, the
state with lowest energy is just the corresponding harmonic oscillator ground
state. For $a_d^{\widetilde 2,0}/l_0 = 69$, the effective potential exhibits two
minima at $r_x = \pm r_{x, 0} \neq 0$. In this situation, it is energetically
advantageous for the molecules to ``pay the price'' of climbing up to the first
excited state in the harmonic potential and thus effectively increase their
relative distance, since this allows them to reduce their total energy due to
the contribution from the dipolar interaction in
Eq.~\eqref{eq:V-eff}. Evidently, the formation of a bound state can be
suppressed by increasing the strength of the harmonic confinement.

The above simplified model does not take symmetry requirements on the
wavefunction for two identical fermions into account. To discuss this point, we
return to the full SE~\eqref{eq:two-body-SE}. We note that a Hamiltonian which
describes the motion of identical particles has to be symmetric under the
exchange $\boldsymbol{\rho}_1 \leftrightarrow \boldsymbol{\rho}_2$ of the
coordinates of the particles. The Hamiltonian in Eq.~\eqref{eq:two-body-SE}
obeys an even stronger symmetry: It is symmetric under the exchange of both only
the $x$ coordinates, $x_1 \leftrightarrow x_2$, and only the $y$ coordinates,
$y_1 \leftrightarrow y_2$. Therefore, also its eigenfunctions have definite
parity under these operations, i.e.,
$\psi(x_1, x_2, r_y) = \pm \psi(x_2, x_1, r_y)$ and
$\psi(x_1, x_2, r_y) = \pm \psi(x_1, x_2, -r_y)$. Overall, the two-body wave
function for identical fermions has to be antisymmetric,
$\psi(x_1, x_2, r_y) = - \psi(x_2, x_1, -r_y)$. Numerically, we find that the
bound state wave function is antisymmetric with respect to
$x_1 \leftrightarrow x_2$, and symmetric under $y_1 \leftrightarrow y_2$ (or,
equivalently, $r_y \to - r_y$). Finally, we note that fermionic statistics imply
that the probability of a close encounter of the molecules is strongly
suppressed, i.e., $\psi(x_1, x_2, r_y) \to 0$ for $x_1 \to x_2$ and $r_y \to 0$.
In comparison to bosonic molecules, this enhances the stability of fermionic
molecules against chemical reactions~\cite{Micheli2010}. Apart from reduced
stability, bound states also occur for pairs of bosonic molecules in the 2LS
configuration. However, for bosons we expect an increased threshold value of the
dipolar length. This expectation is based on the simplified model described
above and on symmetry arguments: The simplified model suggests that in order to
form a bound state, the relative motion of two molecules in the $x$ direction
has to populate excited harmonic oscillator states. For fermions, antisymmetry
of the total wavefunction implies that excited states with odd harmonic
oscillator quantum numbers are admissable, and the lowest lying state that is
compatible with this requirement is the first excited state. However, this state
is excluded for bosons by symmetry. The necessity to invest at least two
harmonic oscillator excitation quanta results in a higher threshold to form a
bound state.

\subsection{Experimental parameters}
\label{sec:exper-param}

\begin{table}[t]
  \centering
  \begin{tabular}{l | c | c | c | c | c | c | c|c}
{} & $\frac{m}{\mathrm{u}}$ & $\frac{\mathcal{D}}{\mathrm{Debye}}$ & $\frac{\hbar B}{h\mathrm{GHz}}$\ & $\frac{\epsilon_0}{\left( \mathrm{kV}/\mathrm{cm}\right)}$ & 
$\frac{l_0}{\mathrm{nm}}$ &$\frac{a^{\widetilde 2,0}_d}{l_0}$ & $\frac{|E_B|}{h\mathrm{kHz}}$ &  $\frac{\Omega}{\mathrm{kHz}}$\\ 
\hline
KRb\textsuperscript{\cite{de2019degenerate}}   & $127$ & $0.57$ & $ 1.11$ & $ 28.9 $  & 61&$1.2$ & 0&$ 86$ \\ %\hline
Toy & $100$ & $1.00$ & $1$ & $14.9 $ & 54& $3.1$ & 0&$135$ \\ %\hline
NaK\textsuperscript{\cite{Park2015, Voges2019}}& $63$ & $2.72$ & $2.83$ & $15.5 $ &  46& $17$ & 0&$307$ \\ %\hline
LiRb & $91$ & $3.99$ & $7.61$ & $28.4$ & 36& $69$ & 66 & $351$ \\ %\hline
LiCs  & $139$ & $5.39$ & $6.54$ & $18.1$ & 28& $241$ & $564^*$ & $374$ 
%NaRb~\cite{Guo2016} & $110$ & $3.2$ & $2.08$ & $9.7$ & $60$ & $360$ \\ \hline
%RbCs~\cite{Molony2014} & $220$ & $1.17$ & $0.49$ & $6.2$ & $15 $ & $146 $ \\ \hline
%NaCs~\cite{Liu2018}  & $156$ & $4.63$ & $1.74$ & $5.6$ & $226$ & $410$
\end{tabular}
  \caption{A list of relevant parameters for different fermionic polar molecules \cite{Zuchowski2013, Aymar2005}. For all  molecular species, the offset field is chosen as $\epsilon_0 =7.5 \,\hbar B/\mathcal{D}$ and the gradient field is fixed to $\epsilon' = 3.5\,(\mathrm{kV}/\mathrm{cm})/\mathrm{mm}$ and $l_0 = 1.5\, l_\perp$ holds. The asterisk ${}^*$ marks binding energies which exceed the single-particle level spacing in the potential well along the confined $z$-axis, $\lvert E_B \rvert>\hbar^2/(2ml_\perp^2) $, and thus go beyond the range of validity of our analysis which assumes that the molecules populate only the lowest-energy single-particle state.}
  \label{tab:parameters}
\end{table}

In this section, we discuss the optimal choice of experimental parameters for the
realization of an interface bound state. From a single-particle point of view we
require the decay rate of molecules in the $+$ channel given in
Eq.~\eqref{eq:2LSlimit} to be small, $\Gamma_\mathrm{2LS}/\omega_0\ll 1$, which
is guaranteed for $l_0/s \ll 1$. However, even a moderately small ratio of
$l_0/s = 1/\sqrt{2}$ leads to $\Gamma_\mathrm{2LS}/\omega_0 < 10^{-6}$, such
that the lifetime of molecules in the $+$ channel is well above any
experimentally relevant time scale. The relation $l_0/s = 1/\sqrt{2}$ can be
rearranged as
\begin{equation}
\hbar \Omega = 2 \left[\epsilon' \mathrm{d}_z^{\widetilde 2,0} (1+\delta)\frac{\hbar}{\sqrt{m}} \right]^{2/3},
\end{equation}
such that $\Omega$ is fixed for a given value of $\epsilon'$. This can be used
to determine the harmonic oscillator length in the adiabatic regime as
\begin{equation}
l_0 = \sqrt{2}\left(\frac{\hbar^2}{2m}\frac{1}{\epsilon'\mathrm{d}_z^{\widetilde 2,0}\frac{1+\delta}{2}}\right)^{1/3}.
\end{equation}
As $l_0$ sets the length and energy scale of the problem, the denominator in the
above expression should be as large as possible. In current experiments, the
electric field gradient is limited by the apparatus. Therefore, the difference
between induced dipole moments
$|\mathrm{d}_z^{\widetilde 2,0}(1+\delta)| = |\mathrm{d}_z^{\widetilde
  2,0}-\mathrm{d}_z^{\widetilde 0,0}|$
has to be maximized. This boils down to finding the optimal value of the
electric offset field, which is $\epsilon_0 = 7.5 \, \hbar B/\mathcal{D}$. In
turn, this implies $\mathrm{d}_z^{\widetilde 0,0} = 0.74 \, \mathcal{D}$,
$\mathrm{d}_z^{\widetilde 2,0} = -0.26 \, \mathcal{D}$ and $\delta= 2.81$. As an
example, we consider the LiRb molecule with parameters
$m = 91 \, \mathrm{u}$, $\mathcal{D} = 3.99 \, \mathrm{Debye}$. Further, we
take the electric field gradient to be $\epsilon' = 3.5 \, \mathrm{(kV/cm)/mm}$.
Then, with the optimal value $\epsilon_0 =7.5 \, \hbar B/\mathcal{D}$ of the
offset field, the harmonic oscillator length evaluates to
$l_0 = 36 \, \mathrm{nm}$, which is considerably below optical length
scales. For the same parameters, the dipolar length is given by
$a_d^{\widetilde 2,0}/l_0 = 69$, which is above the threshold value for the
appearance of a bound state. As can be seen in Fig.~\ref{fig:tls}(a), the
corresponding binding energy is $|E_B| = 2\pi\times 66\,\mathrm{kHz}\,\hbar$ for the given parameters. In
comparison, recent experiments with KRb~\cite{de2019degenerate} reached
temperatures which correspond to a much lower energy of
$50\, \mathrm{nK} \, k_{\mathrm{B}} = 2\pi \times 1\, \mathrm{kHz}\,\hbar$.  We
list additional parameters for different species of polar molecules in
Table~\ref{tab:parameters}.

We finally point out that the dipolar length scales as
$a_d^{\widetilde 2,0}/l_0\propto \mathcal{D}^2\left(\mathcal{D}\epsilon'
\right)^{1/3}m^{4/3}$
with a prefactor that depends on the value of the offset field $\epsilon_0$. In
experiments, this dependence on $\epsilon_0$ can be used to tune the dipolar
length by adjusting the value of $\epsilon_0$, and thus cross the threshold for
the formation of a bound state.

\subsection{Adiabatic loading}
\label{sec:adiabatic-loading-2LS}

To conclude this section, we present an experimental protocol to
adiabatically prepare molecules in the $+$ channel. The electric offset field is
kept at a constant value $\epsilon_0$ throughout the protocol, while the
gradient field $\epsilon'$ as well as the Rabi coupling $\Omega$ in the 2LS
Hamiltonian~\eqref{eq:2LSHam} are set to zero
initially. Further, we assume that the molecules are prepared in the rotational
state $\ket{\widetilde 0,0}$, and that they are trapped in an auxiliary optical
potential $V_\mathrm{aux}(x)$, which is switched off at the end of the protocol. 

The first step is to turn on the Rabi coupling in the 2LS
Hamiltonian~\eqref{eq:2LSHam}, where the detunings are chosen such that
$\Delta_{\widetilde{2}, \widetilde{0}} = \Delta_{\widetilde2,0} -
\Delta_{\widetilde0,0} < 0$
and $\lvert \Delta_{\widetilde{2}, \widetilde{0}} \rvert/\Omega \gg 1$, implying
$\ket{+} \approx \ket{\widetilde 0,0}$. Next, the magnitue of the detunings is
adiabatically reduced
($\lvert \partial_t{\Delta}_{\widetilde 2, \widetilde 0} \rvert/\Omega^2 \ll 1$)
to zero.  Under this condition, the molecules remain in the instantaneous
excited eigenstate $\ket{+}$ and at the end of the adiabatic sweep this state is
given by $\ket{+}=( \ket{\widetilde 0,0}+\ket{\widetilde 2,0})/\sqrt{2}$.  The
final state to prepare molecules in $\ket{+}_x$ from
Eq.~\eqref{eq:plus-minus-channel-states} is to adiabatically turn off the
auxiliary optical potential while ramping up the electric gradient field
$\epsilon'$, such that the effective harmonic confinement in the $+$ channel
($E_+^\mathrm{2LS}(x)+V_\mathrm{aux}(x)$) stays approximately constant.  At the
same time, states in the $-$ channel evolve from being trapped to being
antitrapped.  Due to nonadiabatic channel couplings, there are narrow avoided
crossings between excited motional states in the $-$ channel and states in the
$+$ channel. This avoided crossings are exponentially small, see
Appendix~\ref{sec:decay-rate-tls}, and have to be passed nonadiabatically.

\section{Conclusion and Outlook}
\label{sec:conclusions}

To summarize, we presented two approaches for engineering quantum many-body
systems with polar molecules by employing the coupling of rotational states via
external laser or MW fields and electric field gradients. The first approach
builds on a Raman coupled $\Lambda$-systems, where the strong electric field
gradient of a standing laser field leads to the generation of a single particle
potential barrier on the nanoscale. The second scheme generates spatially
modulated electric dipoles and thus dipolar interactions by MW mixing in the
presence of electric DC field gradients.

Nanostructured optical barriers in combination with optical lattice potentials
enable the generation of bilayer systems, where the layers are separated by only
few tens of nanometers. The resulting many-body systems are described by a
Hamiltonian of the form
\begin{equation}
  H=H_{L}+H_{R}+\sum_{{i_L},{j_R}}V_{LR}(\boldsymbol{\rho}_{i_L}-\boldsymbol{\rho}_{j_R}),
\end{equation}
where
$H_{\alpha}=\sum_{i}(-\hbar^{2}/2m)\nabla^2_{i_\alpha}+\sum_{i<j}V_{\alpha\alpha}(\boldsymbol{\rho}_{i_\alpha}-\boldsymbol{\rho}_{j_\alpha})$
is the Hamiltonian which describes the motion and interactions of molecules
within the layer $\alpha = L, R$, $\boldsymbol{\rho}_{i_\alpha}$ the position of
molecule $i$ in the layer $\alpha=L,R$, and $V_{LR}$ describes the interaction
between molecules in different layers. The strong enhancement of $V_{LR}$ due to
the small layer separation makes interesting few- and many-body physics
experimentally accessible for molecules with a dipole moment of less than one
Debye. Examples include the formation of interlayer bound states as discussed in
this work, the BCS to BEC crossover for fermionic molecules~\cite{Zinner2012},
bilayer quantum Hall physics and---if the motion of molecules within each layers
is restricted further to a one-dimensional channel---ladder physics (see, for
example, Ref.~\cite{baranov2012condensed} and references therein). 
We note that this scheme can be applied to atoms carrying magnetic dipole moments as originally proposed in Ref.~\cite{lkacki2016nanoscale}.

The second, all electric approach enables the realization of many-body systems
in which the induced dipole moment $\mathrm{d}_z^+(z)$ changes sign on a spatial
scale of tens of nanometers. The corresponding many-body Hamiltonian is given by
\begin{equation}
  H = \sum_{i}\left[-\frac{\hbar^{2}}{2m}\nabla^2_{i} +
    E_{+}^{\mathrm{2LS}}(x_{i})\right] + \sum_{i<j}
  \frac{\mathrm{d}^+_{z}(x_{i})
    \mathrm{d}^+_{z}(x_{j})}{|\boldsymbol{\rho}_{i} -
    \boldsymbol{\rho}_{j}|^{3}},
\end{equation}
where $E_{+}^{\mathrm{2LS}}(x)$ is an effective single-particle potential, whose
explicit form is stated in Eq.~\eqref{eq:Epm}. At $x = 0$, the induced dipole
moment vanishes. Two molecules which are located on opposite sides of this
interface at $x = 0$ interact attractively and can form bound states, while
molecules on the same side of the interface always repel each other. The
observation of these bound states under current experimental conditions is
possible for molecules with dipole moments of a few Debye. In a system of
fermionic molecules, the formation of interface bound state can cause a
transition from fermionic to bosonic behavior, and it is an interesting question
for further studies which quantum phases can be realized in such systems. We
finally note that this all electric scheme can be applied with molecules for
which the complexity of the level structure makes the optical manipulation with
laser light challenging. Moreover, this technique can also be adapted to atoms
or molecules with magnetic dipole moments by coupling to an external magnetic field gradient~\cite{Hofferberth2006, Perrin2017}, 
which requires gradients on the order of thousands of G/mm~\cite{note:magnetic}.

In summary, the promise of enhanced energy scales and novel interaction terms
modulated on spatial scales of tens of nanometers provides an interesting new
avenue for many-body physics, including BCS--BEC or fractional quantum Hall
phases. 

{\em Acknowledgments} --The authors thank M. Lacki, D. Petter and M. Mark for
discussions.  P.Z. thanks JILA for hospitality as Visiting Fellow in September
2018, where part of this work was done.  Work at Innsbruck is supported by the
European Union program Horizon 2020 under Grants Agreement No.~817482 (PASQuanS)
and No.~731473 (QuantERA via QTFLAG), the US Air Force Office of Scientific
Research (AFOSR) via IOE Grant No.~FA9550-19-1-7044 LASCEM, by the Simons
Collaboration on Ultra-Quantum Matter, which is a grant from the Simons
Foundation (651440, P.Z.), by a joint-project grant from the FWF (I 4426,
RSF/Russia 2019), and by the Institut f\"ur Quanteninformation. Work at Boulder
is supported by the Army Research Office (ARO) Grant W911NF-19-1-0210, the
National Science Foundation (NSF) Grant No. PHY-1820885, JILA-NSF Grant
No. PFC-1734006, and the National Institute of Standards and Technology (NIST).

\appendix

\section{Validity of the Born-Oppenheimer approximation}
\label{sec:decayBO}

Due to the large separation of typical energy scales between the internal
rotational and the external motional degrees of freedom of polar molecules, the
internal state of a molecule follows its motion essentially adiabatically, as discussed in the main text 
Secs.~\ref{sec:LambdaSystem} and~\ref{sec:2LS}. In
particular, the internal state changes according to the spatial variation of
applied electric fields. Deviations from such fully adiabatic dynamics are
discussed in this Appendix. First, we discuss the transformation of the
Hamiltonian into the BO basis and classify diagonal and off-diagonal
corrections. Then, we study the impact of diagonal nonadiabatic corrections on
wave functions in the BO channels of interest, which is the zero-energy channel
in the $\Lambda$-system and the $+$ channel in the 2LS. Last, we quantify the
decay of states in these BO channels due to nonadiabatic channel couplings. We
estimate the corresponding decay rates, which are given in
Eqs.~\eqref{eq:Lambda-decay-rate} and~\eqref{eq:2LSlimit} in the main text,
using Fermi's golden rule, and show that they are exponentially suppressed.

\subsection{Basis transformation}
\label{sec:basistrans}
For the sake of completeness we summarize how to perform the BO approximation~\cite{burrows2017nonadiabatic,
  kazantsev1990mechanical, dum1996gauge, dutta1999tunneling, ruseckas2005non} for the $\Lambda$-system~\cite{lkacki2016nanoscale,jendrzejewski2016subwavelength} discussed in Sec.~\ref{sec:Lambda}
  and apply these ideas to the TLS from Sec.~\ref{sec:ssp2LS}. As a first step the system Hamiltonian has to be
transformed into the spatially dependent BO basis. In particular, we consider
the Hamiltonians in Eqs.~\eqref{eq:H-Lambda} and~\eqref{eq:H-2LS}, which
describe the dynamics of a single molecule in the $\Lambda$-system and 2LS
setups, respectively. By introducing an index
$\alpha\in\{\Lambda,\,\mathrm{2LS}\}$, these Hamiltonians can be brought to the
common form
\begin{equation}
H_\alpha = \frac{p_z^2}{2m}+\delta_{\Lambda, \alpha}\,V_L(z) +H^0_\alpha(z).
\label{eq:Halpha}
\end{equation}
The Kronecker delta $\delta_{\Lambda, \alpha}$ ensures that the external
confinement $V_L(z)$ is only present in the $\Lambda$-system. We further note
that in the 2LS, the $z$ coordinate has to be replaced by $x$. The adiabatic or
BO basis is formed by the eigenvectors of $H^0_\alpha(z)$, which satisfy
$H^0_\alpha(z)\,\ket{\sigma_\alpha}_z, =
E^\alpha_\sigma(z)\ket{\sigma_\alpha}_z$,
and depend on the position $z$. For the $\Lambda$-system, where
$\sigma = 0,\pm$, the eigenvectors and energies are given in
Eqs.~\eqref{eq:brightEnergy} and~\eqref{eq:zero-energy-BO-state}; for the 2LS,
$\sigma = \pm$, and the eigenvectors and energies are given in
Eqs.~\eqref{eq:Epm} and~\eqref{eq:plus-minus-channel-states}. To transform the
Hamiltonian~\eqref{eq:Halpha} to the BO basis, we expand the state of a molecule
as $\ket{\Psi} = \int \mathrm{d} z \sum_{\sigma} \Psi_{\sigma}(z) \ket{z}\otimes \ket{\sigma_\alpha}_z$,
where $\ket{z}$ is an eigenstate of the position operator. The effective
Hamiltonian $H_{\alpha}^{\mathrm{ad}}$ for the wave functions
$\Psi_{\sigma}(z)$ in the adiabatic basis can be obtained from
Eq.~\eqref{eq:Halpha} by shifting the momentum operator according to
$p_z \rightarrow p_z - A_\alpha(z)$, where $A_\alpha(z)_{\sigma,\sigma'} = - {}_z\bra{\sigma_\alpha}p_z\ket{\sigma'_\alpha}_z$
can be interpreted as a gauge potential,
\begin{equation}
  \label{eq:H-alpha-adiabatic-basis}
  H_{\alpha}^{\mathrm{ad}} = \frac{p_z^2 - \{p_z, A_\alpha(z)\} + A_\alpha^2(z)}{2m} + D_\alpha(z),
\end{equation}
with BO potentials 
\begin{equation*}
\begin{split}
D_\Lambda(z) &= V_L(z) + \mathrm{diag}\left[E^\Lambda_+(z),\,0,\,E^\Lambda_-(z)\right],\\
D_\mathrm{2LS}(z) &=  \mathrm{diag}\left[E^\mathrm{2LS}_+(z),\,E^\mathrm{2LS}_-(z)\right].
\end{split}
\end{equation*}
The gauge potential for the $\Lambda$-system can be found in 
the supplementary material of~\cite{lkacki2016nanoscale} or in the main text of Ref.\cite{jendrzejewski2016subwavelength} and is given by 
\begin{equation}
\label{eq:ALambda}
  A_\Lambda(z)=\frac{i\hbar}{l\sqrt{2}}\frac{1}{1+(z/l)^{2}}
  \begin{pmatrix}
    0 & 1 & 0 \\
    -1 & 0 & 1 \\
    0 & -1 & 0
  \end{pmatrix}
\end{equation}
for $\Delta = 0$ and $l\ll \lambda_c$.
For the 2LS, the gauge potential reads
\begin{equation}  
  A_\mathrm{2LS}(z) = \frac{i\hbar}{2s}\frac{1}{1+(z/s)^2}
  \begin{pmatrix}
    0 & -1 \\
    1 & 0
  \end{pmatrix}.
\end{equation}
The contributions from $A_\alpha(z)$ are twofold: Since $A_\alpha(z)$ is purely
off-diagonal in the adiabatic basis spanned by the states
$\ket{\sigma_\alpha}_z$, the term $\{p_z , A_\alpha(z)\}$ describes nonadiabatic
channel couplings; On the other hand, $A_\alpha(z)^2$ contains for the
$\Lambda$-system both off-diagonal channel couplings and diagonal contributions,
and for the 2LS only diagonal contributions.  Diagonal contributions give rise
to repulsive potential barriers. If nonadiabatic channel couplings are
negligible, the Hamiltonian for a BO channel $\sigma$ is given by
\begin{equation}
H_{\alpha,\sigma\sigma}^\mathrm{ad} = \frac{p^2_z}{2m} + D_\alpha(z)_{\sigma\sigma} + V_\mathrm{na}(z),
\end{equation}
where $V_\mathrm{na}(z) = {(A_{\alpha})^2{}_{\sigma\sigma}}/{2m}$
is the nonadiabatic potential barrier. For $\sigma = 0$ and $\sigma = +$ we
obtain the Hamiltonians in Eqs.~\eqref{eq:zero-energy-BO-channel}
and~\eqref{eq:+-BO-channel} for the $\Lambda$-system and 2LS, respectively. In
the following sections, we specify the conditions under which nonadiabatic
channel couplings are negligible.
%The potential barrier introduced in \eqref{eq:Lambda-na-potential-barrier} and \eqref{eq:2LS-na-potential-barrier} for the $\Lambda$-system and 2LS, respectively, are diagonal nonadiabatic corrections from the term $A_\alpha^2
%

\subsection{Harmonic confinement in the presence of the barrier}
%Reduction of the wavefunction under the barrier
\label{sec:diag-nonad-corr}

The dynamics of molecules in the zero-energy channel in the $\Lambda$-system
and in the $+$
channel in the 2LS is described by the Hamiltonians in
Eqs.~\eqref{eq:zero-energy-BO-channel} and~\eqref{eq:+-BO-channel},
respectively. Both Hamiltonians take the same form if we treat for the 2LS the
effective BO potential $E_+(x)$
in the $+$
channel in the harmonic approximation~\eqref{eq:harmonic-approx-+-channel} and
neglect the energy shift $\Delta
E$. Then, the generic Hamiltonian in the BO channel of interest reads
\begin{equation}
  \label{eq:diagonal-nonadiabatic-H}
  H = -\frac{\hbar^{2}}{2m}\frac{d^{2}}{dz^{2}} + \frac{1}{2} m \omega_L^2 z^2
  +\frac{\hbar^{2}}{2m l^2} \frac{1 - \lambda^2}{(1+z^2/l^{2})^{2}},
\end{equation}
where the last term corresponds to the diagonal nonadiabatic correction. The
parameter $\lambda$ takes the value $\lambda = 0$ for the $\Lambda$-system. Instead, for the 2LS, $\lambda = \sqrt{3}/2$, and one should replace the
$z$ coordinate by $x$, $\omega_L$ by $\omega_0$, and $l$ by $s$.

In the 2LS, we are interested in the limit $l_0/s \ll 1$, where
$l_0 = \sqrt{\hbar/(m\omega_0)}$ is the characteristic length scale associated
with the frequency $\omega_0$. This inequality implies that the spacing of
energy levels in the harmonic potential in
Eq.~\eqref{eq:diagonal-nonadiabatic-H} is much larger than the strength of the
nonadiabatic potential, $\hbar \omega_0 \gg \hbar^2/(2 m s^2)$. Under this
condition, the nonadiabatic potential is indeed only a negligibly small
perturbation.

In contrast, in the $\Lambda$-system, we require the inverted relation
$a_L \gg l$, where $a_L$ is the length scale associated with $\omega_L$. This
condition guarantees that the nonadiabatic potential barrier is much narrower
than a single well in an optical lattice potential as illustrated in
Fig.~\ref{fig:leak1}(a). As we show in the following, this condition also
implies that low-energy eigenstates of the
Hamiltonian~\eqref{eq:diagonal-nonadiabatic-H} are strongly suppressed inside
the barrier. We thus consider the eigenvalue equation of the
Hamiltonian~\eqref{eq:diagonal-nonadiabatic-H},
\begin{equation}
  H \psi_{\lambda}(z) = E \psi_{\lambda}(z).  
\end{equation}
It is convenient to introduce the new independent variable $\overline{z}=z/l$
and rewrite this equation as 
\begin{equation}
  \left[-\frac{d^{2}}{d\overline{z}^{2}} + \frac{l^4}{a_L^4} \overline{z}^2  + \frac{1 -
      \lambda^2}{(1+\overline{z}^{2})^{2}}\right]\psi_\lambda(\overline{z}) =
  \frac{2ml^{2}}{\hbar^{2}}E\psi_\lambda(\overline{z}).
\label{eq:Schroedinger dimensionless}
\end{equation}
For energies smaller than the height of the barrier, $E\ll\hbar^{2}/(2ml^{2})$,
and inside the barrier, $\left|\overline{z}\right|\lesssim 1$, one can neglect
both the harmonic potential and the right-hand side of this equation, such that
the behavior of low-energy wave functions becomes energy independent. To
establish this universal behavior, we consider the zero-energy solution
$\varphi_\lambda(\overline{z})$ of the SE~\eqref{eq:Schroedinger dimensionless},
\begin{equation}
\left[\frac{d^{2}}{d\overline{z}^{2}}- \frac{1 - \lambda^2}{(1+\overline{z}^{2})^{2}}\right]\varphi_\lambda(\overline z)=0.
\label{eq:dimensionless equation}
\end{equation}
We can find a general solution of this equation by introducing the new
independent variable $y=\arctan(\overline z)$, such that the equation takes the
form
\begin{equation}
\left[ \frac{d^{2}}{dy^{2}}-2\tan(y)\frac{d}{dy}-\left( 1- \lambda^2\right)
\right] \varphi_\lambda(\overline z)=0.
\end{equation}
After introducing the new unknown function $g_\lambda(y)=\varphi_\lambda(y)/\cos(y)$,
we finally obtain
\begin{equation}
\left( \frac{d^{2}}{dy^{2}}-\lambda^{2} \right) g_\lambda(y)=0.
\end{equation}
This equation can easily be solved
and the general solution of Eq. \eqref{eq:dimensionless equation}
then reads
\begin{multline}
  \varphi_\lambda(\overline{z})=\sqrt{1+\overline{z}^{2}} \left\{
    A\cos[\lambda\arctan(\overline{z})] \vphantom{\frac{B}{\lambda}} \right. \\
  \left. +\frac{B}{\lambda}
    \sin[\lambda\arctan(\overline{z})]\right\}, \label{eq:general solution}
\end{multline}
where $A$ and $B$ are unknown constants.
%For the $\Lambda$-system we have $\lambda=0$, therefore the universal solution under the barrier reduces to 
%\begin{equation}
%\varphi_0(\overline z) = \sqrt{1+\overline{z}^{2}}\bigg\{A+B~\mathrm{arctan}(\overline z)\bigg\}.
%\end{equation}
The general solution $\varphi_\lambda(\overline z)$ can be used to formulate the
effective ``boundary conditions'' which connect the wave function and its
derivative on different sites of the barrier. These conditions are applicable
for wave functions which change on a scale that is much larger than the width of
the barrier. In other words, the corresponding eigenenergies are much smaller
than the height of the barrier. The ``boundary conditions'' are given by
\begin{equation}
\begin{split}
\label{eq:function difference}
\psi_\lambda'(0^+)+\psi_\lambda'(0^-) & =
s\lambda\tan\left(\frac{\pi\lambda}{2}\right)\left[\psi'_\lambda(0^+)-
  \psi'_\lambda(0^-)\right], \\
\psi_\lambda(0^+)-\psi_\lambda(0^-) &
=-s\lambda \cot \! \left( \frac{\pi\lambda}{2} \right) \left[
  \psi_\lambda'(0^+)+\psi_\lambda'(0^-) \right],
\end{split}
\end{equation}
where $\psi_\lambda(0^\pm)$ and $\psi'_\lambda(0^\pm)$ are the values of the
wave function and its derivative, respectively, on the right and left side of
the barrier.

To derive the above conditions, let us consider the asymptotics of
$\varphi_\lambda(\overline z)$ for $\overline{z}\gg1$,
\begin{multline}
  \varphi_\lambda(\overline z)\approx\left[A\lambda \sin \! \left(
      \frac{\pi \lambda}{2} \right)-B \cos \! \left( \frac{\pi \lambda}{2}
    \right)\right]\\
 + \left[A \cos \! \left( \frac{\pi
        \lambda}{2} \right)+\frac{B}{\lambda} \sin \! \left( \frac{\pi
        \lambda}{2} \right)\right]\overline{z},
\end{multline}
and for $\overline{z}\ll-1$
\begin{multline}
  \varphi_\lambda(\overline{z})\approx\left[A\lambda \sin \! \left(
      \frac{\pi \lambda}{2} \right)+B \cos \! \left( \frac{\pi \lambda}{2}
    \right)\right]\\+\left[-A \cos \! \left( \frac{\pi
        \lambda}{2} \right)+\frac{B}{\lambda} \sin \! \left( \frac{\pi
        \lambda}{2} \right)\right]\overline{z}.
\end{multline}
These asymptotics have to be matched with the wave function and its
derivatives for $\overline{z}\sim\pm1$, $\psi(\overline{z})\approx\psi_\lambda(0^\pm)+\psi'_\lambda(0^\pm)\overline{z}$.
This gives
\begin{equation}
\begin{split}
\psi_\lambda(0^\pm) & =A\lambda \sin \! \left( \frac{\pi \lambda}{2} \right)\mp B \cos \! \left( \frac{\pi \lambda}{2} \right),\\
\psi'_\lambda(0^\pm) & = \pm A \cos \! \left( \frac{\pi \lambda}{2} \right)+\frac{B}{\lambda} \sin \! \left( \frac{\pi \lambda}{2} \right).
\end{split}
\end{equation}
After excluding the unknown constants $A$ and $B$, and restoring the original
units, we arrive at the conditions presented in Eq.~\eqref{eq:function
  difference}.

The boundary conditions in Eq.~\eqref{eq:function difference} allow us to
estimate the suppression of the wave function in the barrier region. For this
purpose we restore the harmonic potential in
Eq.~\eqref{eq:diagonal-nonadiabatic-H} characterized by the frequency $\omega_L$
and the harmonic length $a_{L}$, which we assume to be much larger than the
width of the barrier $l$, $a_{L}\gg l$. The eigenenergies and eigenfunctions of
this problem can be obtained by matching the solution
\begin{multline}
  y_{+}(\widetilde z)=C_{+}\exp(-\widetilde z^{2}/2)\left[
    \frac{1}{\Gamma(\frac{1-\nu}{2})} \Phi \! \left(-\frac{\nu}{2};\frac{1}{2};\widetilde
      z^{2}\right) \right. \\ \left. -\frac{2\widetilde
      z}{\Gamma(-\frac{\nu}{2})} \Phi \! \left(\frac{1-\nu}{2};\frac{3}{2};
      \widetilde z^{2}\right)\right]
\end{multline}
which decays exponentially for $\widetilde z\to+\infty$ where
$\widetilde z=z/a_L$, with the solution
\begin{multline}
  y_{-}(\widetilde z)=C_{-}\exp(-\widetilde z^{2}/2)\left[
    \frac{1}{\Gamma(\frac{1-\nu}{2})} \Phi \! \left(-\frac{\nu}{2};\frac{1}{2};\widetilde
      z^{2}\right) \right. \\ \left. +\frac{2\widetilde
      z}{\Gamma(-\frac{\nu}{2})} \Phi \! \left(\frac{1-\nu}{2};\frac{3}{2};\widetilde
      z^{2}\right)\right]
\end{multline}
which decays exponentially for $\widetilde z\to-\infty$, in the region of the
barrier. Here, $\Gamma(\widetilde z)$ is the Gamma function,
$\Phi(a;b;\widetilde z)$ the degenerate hypergeometric function, and
$\nu=2E/\hbar\omega_L$ the energy measured in units of the harmonic oscillator
spacing. For low-energy eigenstates, $E\ll\hbar^{2}/(2ml^{2})$, we can use the
boundary conditions~\eqref{eq:function difference}, which give the following
equations:
\begin{align}
  \left[\frac{1}{\Gamma(\frac{1-\nu}{2})}+\frac{l}{a_L}\frac{2}{\Gamma(-\frac{\nu}{2})}\lambda \tan \! \left( \frac{\pi \lambda}{2} \right)\right](C_{+}+C_{-}) & =0,\label{eq:for odd solutions}\\
  \left[\frac{1}{\Gamma(\frac{1-\nu}{2})}\frac{1}{\lambda} \tan \! \left( \frac{\pi \lambda}{2} \right)-\frac{l}{a_L}\frac{2}{\Gamma(-\frac{\nu}{2})}\right](C_{+}-C_{-}) & =0.\label{eq:for even solutions}
\end{align}
These equations determine the eigenergies $\nu$. The eigenenergies of states
which are symmetric with respect to $\widetilde z\to- \widetilde z$ fulfill
\begin{equation}
\frac{1}{\Gamma(\frac{1-\nu}{2})}+\frac{l}{a_L}\frac{2}{\Gamma(-\frac{\nu}{2})}\lambda \tan \! \left( \frac{\pi \lambda}{2} \right)=0,
\end{equation}
where $C_{+}=C_{-}$ follows from Eq.~\eqref{eq:for odd solutions}. The corresponding eigenfunctions for $\left|z\right|\gtrsim l$
have the form
\begin{multline}
  y_{\mathrm{sym}}(z)\sim\exp(-z^{2}/2a_L^{2})\left[ \frac{l}{a_L}\lambda \tan
    \!  \left( \frac{\pi \lambda}{2} \right) \right. \\ \left. \times \Phi \!
    \left(-\frac{\nu}{2};\frac{1}{2};\frac{z^{2}}{a_L^{2}}\right) +\frac{\left|z\right|}{a_L} \Phi \!
    \left(\frac{1-\nu}{2};\frac{3}{2};\frac{z^{2}}{a_L^{2}}\right)\right].
\label{eq:even solution}
\end{multline}
The eigenenergies of antisymmetric solutions, for which $C_{+}=-C_{-}$ results
from Eq.~\eqref{eq:for even solutions}, satisfy the equation
\begin{equation}
\frac{1}{\Gamma(\frac{1-\nu}{2})}\frac{1}{\lambda} \tan \! \left( \frac{\pi \lambda}{2} \right)-\frac{l}{a_L}\frac{2}{\Gamma(-\frac{\nu}{2})}=0,
\end{equation}
and the wave functions for $\left|z\right|\gtrsim l$ are
\begin{multline}
  y_{\mathrm{asym}}(z)\sim\exp(-z^{2}/2a_L^{2})\left[
    \mathrm{sign}(z)\frac{l}{a_L}\frac{\lambda}{ \tan \! \left( \frac{\pi
          \lambda}{2} \right)} \right. \\ \left. \times \Phi \!
    \left(-\frac{\nu}{2};\frac{1}{2};\frac{z^{2}}{a_L^{2}}\right)+\frac{z}{a_L}
    \Phi \!
    \left(\frac{1-\nu}{2};\frac{3}{2};\frac{z^{2}}{a_{L}^{2}}\right)\right]
  .\label{eq:odd solution}
\end{multline}
From Eqs.~\eqref{eq:even solution} and~\eqref{eq:odd solution}, we see that as
stated above the wave function in the region of the barrier,
$\left|z\right|\lesssim l$, is reduced by a factor $l/a_L\ll1$, as compared to
its typical values outside the barrier for $\left|z\right|\gtrsim l$.

\subsection{Nonadiabatic channel couplings}
\label{sec:nonad-chann-coupl}
We now turn to nonadiabatic channel couplings, which correspond to the terms
$C_{\mathrm{na}} = \left[- \{ p_z, A(z) \}+A^2(z)\right]/(2m)$ and
$- \{ p_x, A(x) \}/(2m)$ in
Eq.~\eqref{eq:H-alpha-adiabatic-basis} for the $\Lambda$-system and 2LS,
respectively. In the $\Lambda$-system, we are interested in the zero-energy BO channel. 
The external potential confines the zero energy BO channel, whereas the effective potential in the $-$ channel
does not confine states with energy bigger than $0$. Therefore, the $-$ channel hosts a continuum of
scattering states, and the nonadiabatic channel coupling
$\left[- \{ p_z, A(z) \}+A^2(z)\right]/(2m)$ induces decay of states in the zero-energy BO channel into the
continuum in the $-$ channel. Below, we estimate the corresponding decay rate
using Fermi's golden rule. In the 2LS, we are interested in the
$+$ channel, for which the effective potential is given by the BO potential in Eq.~\eqref{eq:+-BO-channel}. 
Decay occurs again to the open $-$ channel.

\subsubsection{DOS in the open channel}

According to Fermi's golden rule, the rate of decay of a given initial state to
an open channel hosting a continuum of final states is determined by the product
of the transition matrix element between the initial and final states, and the
DOS in the open channel. In this section, we derive the DOS in the open channel,
which corresponds to the $-$ BO channel in both the $\Lambda$-system and the
2LS.  We neglect nonadibatic corrections to the effective potential in the $-$
channel% , which is thus determined by Eqs.~\eqref{eq:Elambda} and~\eqref{eq:Epm}
% for the $\Lambda$-system and 2LS, respectively
. Moreover, for the $\Lambda$-sytem we are only interested in a spatial region
of extension $|z-z_0|\lesssim a_L\ll 2\pi/k_c$, thus we expand
$\sin(k_cz)\approx k_c z$, and for simplicity we consider for the 2LS the
symmetric case $\delta = 1$. Then, for both systems, the effective BO potential
in the $-$ channel can be written as
$-U(z)=-\hbar\sqrt{(\Omega/2)^{2}+(\Delta'z)^{2}}$, where we use the following
identification table:
\begin{center}
\begin{tabular}{ l | c | c }
 ~ & $\Omega$ & $\Delta'$ \\ 
 \hline
 2LS & $~\Omega ~$ & $~\Delta'_{\widetilde 2, 0}~$ \\  
$\Lambda$-system & $~\Omega_p~$ & $~\Omega_ck_c/2$
\end{tabular}
\end{center}
For the 2LS the $z$ coordinate has to be replaced by $x$, and for the
$\Lambda$-system we consider the resonant case with $\Delta = 0$.  Since we are
interested in the resonant decay from energetically higher channels into the $-$
channel, only states with energy $E>-\Omega/2$ are relevant. Such states can be
described in the WKB approximation. To write down the corresponding wave
functions, we assume that the system is contained in a large box of size $2L$,
$\left|x\right|\leq L$. (The size $L$ will disappear from the final result for
the decay rate.) Then, the normalized wave function can be written as
\begin{multline}
  \psi_{E}^{(-)}(z)=\frac{1}{L^{1/4}}\frac{1}{\sqrt{2}}\left[\frac{\Delta'}{E+U(z)}\right]^{1/4}
  \\ \times \sin/\cos\left\{ \frac{1}{\hbar}\int_{0}^{z}dz'\sqrt{2m[E+U(z)]}\right\} ,
\label{eq:Wave function open channel}
\end{multline}
where we neglect terms of order $\Omega/\Delta' L\ll1$. The choice of
$\sin\{\ldots\}$ or $\cos\{\ldots\}$ corresponds to either symmetric or
antisymmetric parity of the wave function, respectively. The corresponding
eigenenergies $E=E_{n}$ can be obtained from the WKB quantization condition
\begin{equation}
  \frac{1}{\hbar}\int_{-L}^{L}dz'\sqrt{2m[E_{n}+U(z)]}=\pi \left( n+\frac{1}{2} \right).
\end{equation}
It follows from this equation that the DOS is given by
\begin{equation}
\frac{dn}{dE}\approx
\frac{2\sqrt{2}}{\pi}\sqrt{\frac{mL}{\hbar^3\Delta'}},
\label{eq:DOS open}
\end{equation}
where we neglect terms vanishing for $L\rightarrow \infty$.

\subsubsection{Decay rate in the Lambda-system}
\label{sec:decay-rate-lambda}

We now derive the decay rate of the zero-energy BO channel in the
$\Lambda$-system for the resonant case $\Delta = 0$. As discussed in
Sec.~\ref{sec:diag-nonad-corr}, for $a_L\gg l$ the nonadiabatic potential
barrier has a strong influence on the wave function in the zero-energy
channel. However, to estimate the decay rate, we neglect the effects of the
barrier on the open-channel wave function $\psi_{E}^{(-)}(x)$. This is
legitimate because the height of the barrier $\hbar^{2}/(2ml^{2})$ is much
smaller than the gap $\Omega_p/2$ between the $-$ channel and the zero-energy
channel. Therefore, we can use Eq.~\eqref{eq:Wave function open channel} with
$E=0$ for the final wave function $\psi_{E}^{(-)}(x)$ in the open channel,
\begin{equation}
\begin{split}\psi_{E=0}^{(-)}(\overline{z}) & \approx\frac{1}{L^{1/4}}\frac{1}{\sqrt{2}}\left[\frac{\Delta'}{-E^\Lambda_{-}(\overline{z})}\right]^{1/4}\\
 & \mathrel{\hphantom{=}}\times\sin/\cos\left\{ \frac{1}{\hbar}\int_{0}^{z}dz'\sqrt{-2mE^\Lambda_{-}(z)}\right\} \\
 & =\frac{1}{(Ll)^{1/4}}\frac{1}{\sqrt{2}}\frac{1}{(1+\overline{z}^{2})^{1/8}}\\
 & \mathrel{\hphantom{=}}\times\sin/\cos\left\{ \kappa\int_{0}^{\overline{z}}d\overline{z}'(1+\overline{z}'^{2})^{1/4}\right\} ,
\end{split}
\label{eq:psi_minus}
\end{equation}
where $\overline{z}=z/l$, $E^\Lambda_{-}(z)$ is defined in
Eq~\eqref{eq:brightEnergy}, and $\kappa$ is the square root of the ratio of the
gap between the channels and the height of the barrier,
$\kappa=[ml^{2}\Omega_{p}/\hbar]^{1/2}$. The nonadiabatic channel coupling
$- \{ p_z, A(z)\}/(2m)$, where $A(z)$ is given by Eq.~\eqref{eq:ALambda} and  
is nonzero only inside the
barrier for $\left|z\right|\lesssim l$, where the behavior of the wave function
in the zero-energy channel is governed by the universal solution in
Eq.~\eqref{eq:general solution} with $\lambda =0$,
\begin{equation}
  \varphi_{0}(\overline{z})=\sqrt{1+\overline{z}^{2}}\left[ A+B \arctan(\overline{z})
  \right].
\end{equation}
The coefficients $A$ and $B$ in the above expression depend on details of the
behavior of the wave function outside the barrier and on the position of the
barrier. We can estimate these coefficients as follows: The typical value of the
wave function localized in a spatial area of size $a_L$ is
$\varphi\sim1/\sqrt{a_L}$; Further, according to our discussion in
Sec.~\ref{sec:diag-nonad-corr}, the wave function within the barrier is reduced
by a factor of $l/a_L$. This yields the estimate $A\sim B\sim l/a_L^{3/2}$. The
nonadiabatic coupling matrix element then reads
\begin{equation}
\begin{split}
    M&=-\int_{-L}^{L}\mathrm{d}z\,\psi_{E=0}^{(-)}(\overline{z})\frac{\hbar^{2}}{ml^{2}\sqrt{2}}\left[\frac{1}{1+\overline{z}^{2}}\frac{d}{d\overline{z}}\right.\\
    &~~~~~~~~~~~~~~~~~~~~~~~~~~~~~~~~~~~~~-\left.\frac{\overline{z}}{(1+\overline{z}^{2})^{2}}\right]\varphi_{0}(\overline{z})\\
    &=-\frac{\hbar^{2}}{ml\sqrt{2}}\int_{-L/l}^{L/l}\mathrm{d}\overline{z}\,\psi_{E=0}^{(-)}(\overline{z})\frac{B}{(1+\overline{z}^{2})^{3/2}}.
    \end{split}
\end{equation}
We note that the term which is proportional to $A$ in
$\varphi_{0}(\overline{z})$ gives a vanishing contribution. A nonvanishing
result can be obtained only if one includes the correction to
$\varphi_{0}(\overline{z})$ of first order in $2ml^{2}E/\hbar^{2}\ll1$ in
Eq.~\eqref{eq:Schroedinger dimensionless}. After substituting the
expression~\eqref{eq:psi_minus} for $\psi_{E=0}^{(-)}(\overline{z})$, we obtain
\begin{equation}
  \label{eq:MatrixElement}
  \begin{split}
M=\frac{\hbar^{2}}{2ml}\frac{B}{\sqrt[4]{Ll}}&\int_{-\infty}^{\infty}\mathrm{d}\overline{z}\frac{1}{(1+\overline{z}^{2})^{13/8}}\\
&\times\cos\left\{ \kappa\int_{0}^{\overline{z}}\mathrm{d}\overline{z}'~(1+\overline{z}'^{2})^{1/4}\right\},
\end{split}
\end{equation}
where, using $L/l\gg1$ and convergence of the integral over $\overline{z}$, we
extend the limits of the integration to infinity. To calculate the integral in
Eq.~\eqref{eq:MatrixElement}, we write it as
\begin{equation}
\label{eq:complex_integral}
\begin{split}
\int_{-\infty}^{\infty}\mathrm{d}\overline{z}\,(1+\overline{z}^{2})^{-13/8}\exp\left\{ i\kappa\int_{0}^{\overline{z}}\mathrm{d}\overline{z}'~(1+\overline{z}'^{2})^{1/4}\right\} ,
\end{split}
\end{equation}
and consider as a contour integral in the complex $\overline{z}$ plane. To
uniquely define the multi-valued integrand, we make two branch cuts on the
imaginary axis: from $i$ to $i\infty$ and from $-i$ to $-i\infty$. We then move
the contour of integration from the real axis to the upper half-plane where the
integrand decays exponentially. The new contour is around the upper branch cut
and consists of three parts: The first one is from $i\infty$ to $i$ over the
left-side of the cut where $\overline{z}=i+y\exp(-i3\pi/2)$ with real
$y\in[\epsilon,\infty]$ and $\epsilon$ infinitesimal and positive. The second
one is around the circle of radius $\epsilon$ around $i$ in the counterclockwise
direction, $\overline{z}=i+\epsilon\exp(i\phi)$ with $\phi\in[-3\pi/2,\pi/2]$,
and the third one is over the right-hand side of the cut,
$\overline{z}=i+y\exp(i\pi/2)$. We note that every integral over individual
parts diverges when $\epsilon\to0$, and only their sum is finite.

On the new integration contour, the integral in the exponent in Eq.~\eqref{eq:complex_integral} can be written as
\begin{equation}
\begin{split}
\int_{0}^{\overline{z}}\mathrm{d}\overline{z}'~(1+\overline{z}'^{2})^{1/4}  =&\int_{0}^{i}\mathrm{d}\overline{z}'~(1+\overline{z}'^{2})^{1/4}\\
+&\int_{i}^{\overline{z}}\mathrm{d}\overline{z}'~(1+\overline{z}'^{2})^{1/4}\\
 \approx& iI+\frac{4}{5}2^{-1/4} e^{i \pi/8} (\overline{z}-i)^{5/4},
\end{split}
\label{eq:phase_function}
\end{equation}
where
\begin{equation}
I=\int_{0}^{1}\mathrm{d}x(1-x^{2})^{1/4}=\frac{\sqrt{2}\pi^{3/2}}{6\Gamma(3/4)}\approx0.874,
\end{equation}
and the second integral is in calculated by keeping only the leading term in the
expansion of $(1+\overline{z}'^{2})^{1/4}$ in powers of $\overline{z}'-i$
because for $\kappa\gg1$ only the region
$|\overline{z}'-i |\lesssim\kappa^{4/5}\ll1$ is important. After substituting
\eqref{eq:phase_function} into \eqref{eq:complex_integral} and expanding
$(1+\overline{z}^{2})^{13/8}$ around $\overline{z}=i$ to leading order, we find
for the integral~\eqref{eq:complex_integral} the result
\begin{equation}
\begin{split} 
& e^{-\kappa I}2^{-5/8}\sqrt{\kappa\frac{4}{5}2^{-1/4}}\intop_{0}^{\infty}\frac{ds}{\sqrt{s}}e^{-s}[\cos(s+\frac{\pi}{8})+\sin(s+\frac{\pi}{8})]\\
& =e^{-\kappa I}2^{-5/8}\sqrt{\kappa\frac{4}{5}2^{-1/4}}\sqrt{\pi\sqrt{2}}=\sqrt{\frac{2\pi}{5}\kappa}e^{-\kappa I}.
\end{split}
\end{equation}
To obtain this result, we first integrate by parts in the $y$-integrals over the
sides of the branch cut (this, in combination with the integral over the circle,
eliminates the divergences for $\epsilon\to0$), then we take the limit
$\epsilon\to0$, and introduce the new integration variable
$s=4\kappa y^{5/4}2^{-1/4}/5$.

With this result, the final expression for the coupling matrix element reads
\begin{equation}
M=\frac{\hbar^{2}}{2ml}\frac{B}{\sqrt[4]{Ll}}\sqrt{\frac{2\pi}{5}\kappa}e^{-\kappa\,I}.
\label{eq:M_Lamba}
\end{equation}
An analogous calculation shows that the channel coupling $A_\Lambda^{2}(z)$ from
Eq.~\eqref{eq:H-alpha-adiabatic-basis} yields a matrix element which is smaller
than $M$ in Eq.~\eqref{eq:M_Lamba} by a factor of order
$\mathcal{O}((l/a_{L})\sqrt{\omega_{L}/\Omega})$ and thus negligible.  With the
matrix element Eq.~\eqref{eq:M_Lamba} and the DOS from Eq.~\eqref{eq:DOS open},
using Fermi's golden rule we obtain for the decay rate
\begin{equation}
\Gamma_\Lambda=\frac{4\pi}{5}\frac{\hbar B^{2}}{ml}\sqrt{\kappa}e^{-2\kappa\,I}.
\label{eq:gamma2}
\end{equation}
We note that the exponential factor in Eq.~\eqref{eq:gamma2} coincides with the results obtained in the absence of harmonic confinement as considered in Ref.~\cite{jendrzejewski2016subwavelength}. 
For the harmonic confinement in the dark-state channel with frequency $\omega_{L}$ one has $B=\beta l/a_{L}^{3/2}$ with $|\beta|\lesssim1$ and $a_{L}=\sqrt{\hbar/m\omega_{L}}$ (for the barrier in the center of the harmonic trap, $\beta=\beta_{0}\approx(4/\pi^{5})^{1/4}=0.34$, such that actually $|\beta|\leq\beta_{0}$). The ratio of the decay rate to the oscillator frequency $\omega_{L}$ then reads 
\begin{equation}
\begin{split}
\frac{\Gamma_{\mathrm{\Lambda }}}{\omega_{L}} & \approx\frac{4\pi}{5}\,\beta^2\left(\frac{l}{a_{L}}\right)^{3/2}\left(\frac{\Omega_{p}}{\omega_{L}}\right)^{1/4}\exp\left(-2\sqrt{\frac{\Omega_{p}}{\omega_{L}}}\frac{l}{a_{L}}~I\right)\\
 & =2.5\,\beta^2\left(\frac{l}{a_{L}}\right)^{3/2}\left(\frac{\Omega_{p}}{\omega_{L}}\right)^{1/4}\exp\left(-1.75\sqrt{\frac{\Omega_{p}}{\omega_{L}}}\frac{l}{a_{L}}\right),
\end{split}
\end{equation}
as written in Eq.~\eqref{eq:Lambda-decay-rate}. We note that the above derivation requires the following hierarchy of scales 
\begin{equation}
  1\gg\frac{l}{a_L}\gg\sqrt{\frac{\omega_L}{\Omega}},
\end{equation}
for which $\Gamma_\Lambda/\omega_L$ is exponentially suppressed.

\subsubsection{Decay rate in the 2LS}
\label{sec:decay-rate-tls}

We proceed to calculate the rate of decay from the $+$ BO channel to the $-$ channel in
the 2LS. For simplicity, we focus on the symmetric case $\delta = 1$, and we
consider the ground state in the harmonic
approximation~\eqref{eq:harmonic-approx-+-channel} to the effective potential in
the $+$ channel as the initial state,%  The harmonic approximation, which is
% discussed in the main text below Eq.~\eqref{eq:H-2LS-adiabatic-basis}, yields
% \begin{equation}
%   \label{eq:harmonic-potential-2LS}
%   E_+(x)\approx \frac{\hbar \Omega}{2}+\frac{m\omega_0^2x^2}{2},
% \end{equation}
% and the corresponding ground state is given by
\begin{equation}
  \phi_{0}(x) = \frac{1}{\sqrt{l_{0}\sqrt{\pi}}} \exp \! \left(-\frac{x^2}{2l_{0}^2} \right),
\end{equation}
where $l_0 = \sqrt{\hbar/(m\omega_0)}$ and we set $x_0 = 0$. The validity of the
harmonic approximation is controlled by the inequality $l_0 \ll s$,
% \begin{align}
% l_0 \ll s,
% \label{eq:appendixInequality}
% \end{align}
where $s$, which is given in Eq.~\eqref{eq:resarea}, denotes the width of the
resonant region. This inequality will be of crucial importance for the following
discussion.

The state $\phi_0(x)$ in the $+$ channel is coupled to the state described by
the wave function $\psi_E^{(-)}(x)$ in Eq.~\eqref{eq:Wave function open channel}
in the $-$ channel by the operator
$C_{\mathrm{na}} = -\left\{p_x,A(x)\right\}/(2m)$, which occurs as a
nonadiabatic channel coupling in Eq.~\eqref{eq:H-alpha-adiabatic-basis}. Using the
aforementioned inequality, this operator can be approximated as
% \begin{equation}
%   \begin{split}
%     C_{\mathrm{na}} & = \frac{\hbar^{2}}{2m}\left[
%       \frac{s}{s^{2}+x^{2}}\frac{d}{dx}-\frac{sx}{(s^{2}+x^{2})^{2}}\right]
%     \left( \ket{+}_x\bra{-}- \mathrm{H.c.} % \ket{-}_x\bra{+}
%     \right) \\ &
%     \approx\frac{\hbar^{2}}{2m}\left[
%       \frac{1}{s}\frac{d}{dx}-\frac{x}{s^{3}}\right] \left(
%     \ket{+}_x\bra{-}- \mathrm{H.c.} % \ket{-}_x\bra{+}
%     \right).
%   \end{split}
% \end{equation}
\begin{equation}
  \begin{split}
    C_{\mathrm{na}} & = \frac{\hbar^{2}}{2m}\left[
      \frac{s}{s^{2}+x^{2}}\frac{d}{dx}-\frac{sx}{(s^{2}+x^{2})^{2}}\right]
    \begin{pmatrix}
    0 & -1 \\
    1 & 0
  \end{pmatrix}
     \\ &
    \approx\frac{\hbar^{2}}{2m}\left[
      \frac{1}{s}\frac{d}{dx}-\frac{x}{s^{3}}\right]
\begin{pmatrix}
    0 & -1 \\
    1 & 0
  \end{pmatrix}.
  \end{split}
\end{equation}
As a result, the coupling matrix element $M$ which enters Fermi's golden rule is
\begin{equation}
M=\frac{\hbar^{2}}{2m}\int_{-L}^{L}\mathrm{d}x~\phi_{0}(x)\left[ \frac{1}{s}\frac{d}{dx}-\frac{x}{s^{3}}\right] \psi_{E}^{(-)}(x)=M_{1}-M_{2}.
\end{equation}
Since $\phi_0(x)$ is a symmetric function of $x$, the matrix element $M$ is
different from zero only if $\psi_E^{(-)}(x)$ is an antisymmetric
function. Therefore, one has to choose the $\sin\{\dots\}$ solution in
Eq.~\eqref{eq:Wave function open channel}. Furthermore, one can easily see that
the contribution from $M_2$ is smaller than that of $M_1$ by a factor which is
$\mathcal{O}((l_{0}/s)^{2})$, and thus
\begin{equation}
M\approx
M_{1}=\frac{\hbar^{2}}{2m}\frac{1}{s}\int_{-L}^{L}dx~\phi_{0}(x)\frac{d}{dx}\psi_{E}^{(-)}(x). 
\end{equation}
The calculation of the above integral can be simplified significantly by using
the inequality $l_{0}\ll s$, which yields the following simplified expression of
$\psi_{E}^{(-)}(x)$ for $\left|x\right|\sim l_{0}$:
\begin{multline}
  \psi_{E}^{(-)}(x)\approx\frac{1}{L^{1/4}}\frac{1}{\sqrt{2}}\left[\frac{\Delta_{\widetilde2,0}'}{E+\hbar\Omega/2}\right]^{1/4}\\
  \times \sin\left\{ \frac{1}{\hbar}x\sqrt{2m[E+\hbar\Omega/2]}\right\} .
\end{multline}
Further, we can expand the oscillatory factor, because the neglected terms are
much smaller than unity if $\omega_0/\Omega\ll 1$. Within this approximation,
we obtain for $E\approx\hbar\Omega/2$ the matrix element
\begin{equation}
  M\approx\frac{1}{L^{1/4}}\frac{\hbar^{2}}{2m}\frac{1}{\sqrt{2l_0}} \left(
    \frac{\pi}{2s} \right)^{1/4}\frac{l_{0}}{s} \widetilde{p} \exp \! \left[
    -\frac{1}{2}(\tilde{p}l_{0})^{2} \right],
\end{equation}
where $\widetilde{p}=\sqrt{4m\Omega}/\hbar$. This result for the matrix element
and the DOS from Eq.~\eqref{eq:DOS open} yield the ratio of the decay rate
$\Gamma_\mathrm{2LS}$ to the oscillator frequency $\omega_0$ given in
Eq.~\eqref{eq:2LSlimit},
\begin{equation}
  \frac{\Gamma_\mathrm{2LS}}{\omega_0} \approx
  2\sqrt{\pi}\frac{l_0}{s}\exp\left[-8\left(\frac{s}{l_0} \right) ^2 \right].
\end{equation}
In particular, we find that the decay rate of the $+$ channel for the 2LS is
determined by the ratio $l_0/s$ and is suppressed exponentially for $l_0 \ll s$.

\section{Interface bound state: Details of numerical analysis}
\label{sec:interface-bound-state-numerics}

The wave function and eigenenergies for the bound state shown in
Fig.~\ref{fig:tls} of the main text Sec.~\ref{sec:2LS-bound-state} are the solutions of a numerical
diagnoalization of Eq.~\eqref{eq:two-body-SE}. For all calculations we use
the linear approximation of the dipole moment,
\begin{equation}
  \mathrm{d}_z^{\mathrm{D}}(x) \approx -\mathrm{d}_z^{\widetilde
    2,0}\frac{4\delta^{3/2}}{(1+\delta)^2}\frac{x}{s},
\end{equation}
as introduced in the main text. It is convenient to express the
SE~\eqref{eq:two-body-SE} from the main text also along the $x$ direction in
relative and COM coordinates, $r_x = x_1 - x_2$ and $R_x = (x_1 + x_2)/2$,
respectively. This leads to
\begin{multline}
\label{eq:disc}
\frac{\hbar^2}{2m l_0^2} \left[-\frac{1}{2}\frac{\partial^2}{\partial
    \widetilde{R}^2_x} - {2}\frac{\partial^2}{\partial \widetilde
    r^2_x}-{2}\frac{\partial^2}{\partial \widetilde r^2_y}+ \left( 2\widetilde
    R_x^2+ \frac{1}{2}\widetilde r_x^2 \right) \right. \\ \left.  +
  \frac{V_{\mathrm{2D}}(\widetilde R_x+\frac{\widetilde r_x}{2},\widetilde
    R_x-\frac{\widetilde r_x}{2}, \widetilde r_y)}{\frac{\hbar^2}{2ml_{0}^2}}
\right] \Psi(\widetilde R_x,\widetilde r_x,\widetilde r_y) \\ =E\Psi(\widetilde
R_x,\widetilde r_x,\widetilde r_y),
\end{multline}
where we measure distances in units of the confinement in the $x$ direction
$l_{0}$, $\widetilde r_x = r_x/l_{0}$, $\widetilde r_y = r_y/l_{0}$, and
$\widetilde{R}_x = R_x/l_{0}$. To obtain the binding energy one has to subtract
from $E$ twice the ground state energy of the noninteracting system, i.e.,
$E_B = E-\hbar\omega_0$. The explicit form of the interaction potential is given
by
\begin{multline}
  \frac{V_{\mathrm{2D}}(\widetilde R_x+\frac{\widetilde r_x}{2},\widetilde
    R_x-\frac{\widetilde r_x}{2}, \widetilde r_y)}{\frac{\hbar^2}{2ml_{0}^2}} \\
  = \frac{1}{\sqrt{2\pi}}\frac{a_d l_{0}}{s^2}\frac{l_0^3}{l_\perp^3} \left(
    \widetilde R_x^2 - \frac{1}{4}\widetilde r_x^2 \right) \exp \!
  \left( \frac{\widetilde\rho^2}{4}\frac{l_0^2}{l_\perp^2} \right) \\ \times \left[ \left(
      2+\widetilde \rho^2\frac{l_0^2}{l_\perp^2} \right) K_0 \left(
      \frac{\widetilde\rho^2}{4}\frac{l_0^2}{l_\perp^2} \right) -\widetilde
    \rho^2\frac{l_0^2}{l_\perp^2}K_1 \left(
      \frac{\widetilde\rho^2}{4}\frac{l_0^2}{l_\perp^2} \right) \right],
\end{multline}
where we define $\widetilde \rho^2 = \widetilde r_x^2 + \widetilde r_y^2$. Due
to the spatial dependence of the dipole moment $\mathrm{d}_z^{\mathrm{D}}(x)$ it
is not possible to separate the relative and COM motion along the $x$ axis. It
is convenient to expand the COM motion in eigenfunctions of the harmonic
potential described by the SE
\begin{equation}
  \frac{\hbar^2}{2ml_{0}^2}\bigg[-\frac{1}{2}\frac{\partial^2}{\partial \widetilde R^2_x}+2\widetilde  R_x^2\bigg]\Phi_n(\widetilde R_x) = \frac{\hbar^2}{2ml_{0}^2}2 \left( n+\frac{1}{2}  \right) \Phi_n(\widetilde R_x).
\end{equation}
That is, we use the ansatz
\begin{equation}
  \Psi(\widetilde R_x, \widetilde r_x,\widetilde r_y) =
  \sum_{n=0}^N\Phi_n(\widetilde R_x) \phi_n \!  \left( \widetilde r_x, \widetilde r_y
   \right) ,
\end{equation}
where we only keep the lowest $N$ eigenfunctions in the above harmonic
potential. We note that the required cutoff $N$ to reach convergent results for
the bound state and bound state energy is usually much smaller than the number
of grid points one needs to obtain comparable results from a brute-fore
discretization in real space. The above ansatz yields the following SE for the amplitudes
$\phi_n(\widetilde r_x,\widetilde r_y)$:
\begin{multline}
  \frac{\hbar^2}{2m l_{0}^2}\sum_{m=0}^N
  \bigg\{\bigg[-{2}\frac{\partial^2}{\partial \widetilde
    r^2_x}-{2}\frac{\partial^2}{\partial \widetilde r^2_y} +
  \frac{1}{2} \widetilde
  r_x^2+2 \left( n+\frac{1}{2} \right) \bigg]\delta_{m,n} \\ +\overline
  V_{2\mathrm{D}}^{m,n} \left( \widetilde r_x, \widetilde r_y
  \right) \bigg\}\phi_m(\widetilde r_x, \widetilde r_y) = \left(E_B-\hbar\omega_0 \right)\phi_n(\widetilde r_x,
  \widetilde r_y).
\end{multline}
Amplitudes $\phi_n(\widetilde{r}_x, \widetilde{r}_y)$ with different $n$ are
coupled by the interaction matrix elements 
% \begin{equation} 
% \begin{split}
% \overline V_{2\mathrm{D}}^{m,n} \left( \widetilde r_x, \widetilde r_y  \right) &=
% \int_{-\infty}^\infty\mathrm{d} \widetilde R_x \Phi_m(\widetilde R_x)
% \frac{V_{\mathrm{2D}}(\widetilde R_x+\frac{\widetilde r_x}{2},\widetilde
%   R_x-\frac{\widetilde r_x}{2}, \widetilde r_y)}{\frac{\hbar^2}{2ml_{0}^2}}
% \Phi_n(\widetilde R_x)\\
% & = \frac{1}{\sqrt{2\pi}}\frac{a_d l_{0}}{s^2}\frac{l_0^3}{l_\perp^3}
%   \exp \! 
%      \left( \frac{\widetilde\rho^2}{4}\frac{l_0^2}{l_\perp^2}  \right) \left[ \left( 2+\widetilde \rho^2\frac{l_0^2}{l_\perp^2}
%        \right) K_0 \left( \frac{\widetilde\rho^2}{4}\frac{l_0^2}{l_\perp^2}
%        \right) -\widetilde \rho^2\frac{l_0^2}{l_\perp^2}K_1 \left( \frac{\widetilde\rho^2}{4}\frac{l_0^2}{l_\perp^2}  \right) 
%     \right]\\
%     &~~~\times \frac{1}{4}\bigg\{\sqrt{(n+1)(n+2)} \delta_{m,n+2}+
%      \left[(2n+1)- \widetilde r_x^2  \right] \delta_{m,n}
%     +\sqrt{n(n-1)}\delta_{m,n-2}
%     \bigg\}.
% \end{split}
% \end{equation}
\begin{multline} 
  \overline V_{2\mathrm{D}}^{m,n} \left( \widetilde r_x, \widetilde r_y  \right) \\
\begin{aligned}
  &= \int_{-\infty}^\infty\mathrm{d} \widetilde R_x \Phi_m(\widetilde R_x)
  \frac{V_{\mathrm{2D}}(\widetilde R_x+\frac{\widetilde r_x}{2},\widetilde
    R_x-\frac{\widetilde r_x}{2}, \widetilde r_y)}{\frac{\hbar^2}{2ml_{0}^2}}
  \Phi_n(\widetilde R_x)\\
  & = \frac{1}{\sqrt{2\pi}}\frac{a_d l_{0}}{s^2}\frac{l_0^3}{l_\perp^3} \exp \!
  \left( \frac{\widetilde\rho^2}{4}\frac{l_0^2}{l_\perp^2} \right) \left[ \left(
      2+\widetilde \rho^2\frac{l_0^2}{l_\perp^2} \right) K_0 \! \left(
      \frac{\widetilde\rho^2}{4}\frac{l_0^2}{l_\perp^2} \right) \right. \\ &
  \mathrel{\hphantom{=}} \left. -\widetilde \rho^2\frac{l_0^2}{l_\perp^2}K_1
    \left( \frac{\widetilde\rho^2}{4}\frac{l_0^2}{l_\perp^2} \right) \right]
  \frac{1}{4}\bigg\{\sqrt{(n+1)(n+2)} \delta_{m,n+2} \\ & \mathrel{\hphantom{=}}
  + \left[(2n+1)- \widetilde r_x^2 \right] \delta_{m,n}
  +\sqrt{n(n-1)}\delta_{m,n-2} \bigg\}.
\end{aligned}
\end{multline}
For the parameters used in Fig.~\ref{fig:tls}, we obtain
convergence for $N < 9$, a numerical box size of $|\widetilde r_{x,y}| < 10$ and a uniform grid of
$600\times600$ for the two relative coordinates.

% \bibliography{bib}

\begin{thebibliography}{69}
\bibitem [{not({\natexlab{b}})}]{note:magnetic}%
  \BibitemOpen
  \href@noop {} {}\bibinfo {howpublished} {We use the fact that the energy of
  an electric dipole of $1$ Debye in the electric field of $1\,\mathrm{kV/cm}$
  is equal to the energy of a magnetic dipole moment of $1$ Bohr magneton in
  the magnetic field of $360$ Gauss} \BibitemShut {NoStop}%
\end{thebibliography}

%merlin.mbs apsrev4-1.bst 2010-07-25 4.21a (PWD, AO, DPC) hacked
%Control: key (0)
%Control: author (8) initials jnrlst
%Control: editor formatted (1) identically to author
%Control: production of article title (-1) disabled
%Control: page (0) single
%Control: year (1) truncated
%Control: production of eprint (0) enabled
%

\end{document}